\documentclass[runningheads]{svmult}

\usepackage{citesort}
\usepackage{makeidx}   
\usepackage{graphicx}  
\usepackage{subeqnar}  
\usepackage{multicol}  
\usepackage{physprbb}  
\makeindex             

\def\CC{{\rm\kern.24em \vrule width.04em height1.46ex depth-.07ex
\kern-.30em C}}

\begin{document}
\title*{Decoherence-Free Subspaces and Subsystems}
\toctitle{}
%
%
\titlerunning{Decoherence-Free Subspaces and Subsystems}
%
\author{Daniel A. Lidar\inst{1}
\and K. Birgitta Whaley\inst{2}}
\authorrunning{Lidar \& Whaley}
%
%
\institute{Chemical Physics Theory Group, Chemistry Department,
University of Toronto, 80 St. George St., Toronto, Ontario M5S 3H6, Canada
\and Department of Chemistry and Kenneth Pitzer Center for Theoretical Chemistry, University of California, Berkeley, California
94720, USA}

\maketitle              

\begin{abstract}
Decoherence is the phenomenon of non-unitary dynamics that arises as a
consequence of coupling between a system and its environment. It has
important harmful implications for quantum information processing, and
various 
solutions to the problem have been proposed. Here we provide a detailed a review of the theory of decoherence-free
subspaces and subsystems, focusing on their usefulness for preservation of
quantum information. 
\end{abstract}

\section{Introduction}

\label{intro}

Recent results indicating that quantum information processing (QIP) is
inherently more powerful than its classical counterpart
\cite{Gruska:book,Kitaev:book,Nielsen:book} have motivated
a resurgence of interest in the problem of decoherence \cite{Giulini:book}. Decoherence is
a consequence of the inevitable coupling of any quantum system
to its environment (or bath), causing information loss from the system to this environment. It was recognized early on that decoherence poses a serious
obstacle to physical realization of quantum information processors
\cite{Landauer:95,Unruh:95}. Here we define decoherence as {\em non-unitary
  dynamics that is a consequence of system-environment coupling}. This
includes (but is not limited to)
both dissipative and dephasing contributions, traditionally known as
$T_1$ and $T_2$ processes, respectively. Dissipation refers to
processes in which the populations of the quantum states are modified by
interactions with the environment, while dephasing refers to processes that
randomize the relative phases of the quantum states. Both are caused by
entanglement of the system with environmental degrees of freedom, leading to
non-unitary system dynamics \cite{note1}. For QIP and other forms of quantum
control, any such interaction that degrades the unitary nature of the quantum evolution is undesirable, since it
causes loss of coherence of the quantum states and hence an inevitable decay
of their interference and entanglement. Interference is crucial for
coherent control schemes \cite{Shapiro:00,Rabitz:00}, while entanglement is believed to be as
important an 
ingredient of the quantum computational speed-up \cite{Ekert:98}. In fact, a sufficiently
decohered quantum computer can be efficiently simulated on a classical
computer \cite{Aharonov:96a}. In this review we provide a detailed introduction to the
theory of decoherence-free subspaces and subsystems (DFSs), which have been conceived
as one of the possible solutions to decoherence in QIP. Space
limitations prevent us from discussing in detail the interesting problem of
performing 
{\em computation} using DFSs. Instead we focus here on the use of DFSs as a way to {\em preserve} delicate quantum information.

\section{Historical Background}
\label{history}
Environment-induced decoherence is a very extensively studied phenomenon in
the quantum theory of measurement, and of the transition from quantum to
classical behavior. Zurek has shown that environmental superselection
results in the establishment of certain special states that show
redundancies in their correlations with the environment, and that may
consequently be essentially unperturbed by this \cite{Zurek:81}. These
states, referred to as ``pointer states'' ,
are defined by a ``predictability sieve'' \cite{Zurek:93}.
While analysis of measurements are complicated by the presence of a
measurement apparatus in addition to the system and environment, the notion
of a special set of states that are defined by some underlying symmetry in
the global physical description is a common element. Another noteworthy early
study employing symmetries in order to reduce the coupling of
subsets of system states to the environment is Alicki's work on limited
thermalization \cite{Alicki:88}. Identifying the power
of such symmetries in the physical Hamiltonian, and systematizing the way to
find such symmetries has been one of the major innovative features of the
recent development of DFS theory and its applications
for quantum computation.

Early discussions of the effects of decoherence on quantum computation
focused on putting conditions on the strength of coupling of the
system-environment coupling and on the duration of quantum gates
\cite{Landauer:95,Unruh:95}. The search for systematic ways to bypass
decoherence in the context of QIP,
based on identification of states that might be immune to certain decohering
interactions, started with observations of 
Palma, Suominen, and Ekert \cite{Palma:96} in a
study of the effects of pure {\em dephasing}, that two qubits possessing identical
interactions with the environment do not decohere. Palma {\it et al}. used the
term ``subdecoherence''\ to describe this
phenomenon, suggested using the corresponding states to form a \textquotedblleft
noiseless''\ encoding into logical qubits, and noted that
the set of states robust against dephasing will depend on the specific form
of qubit-environment coupling. This model was subsequently studied
using a different method by Duan and Guo \cite{Duan:98}, with similar
conclusions and a change of terminology to ``coherence preserving states''. The idea of pairing qubits as a means
for preserving coherence was further generalized by Duan and Guo in \cite{Duan:97PRL}, where it was shown that both collective dephasing
and dissipation could be prevented. However this assumed knowledge of the system-environment
coupling strength. These
early studies were subsequently cast into a general mathematical framework
for DFSs of more general system/environment
interactions by Zanardi and Rasetti, first for the spin-boson model
in \cite{Zanardi:97c}, where the important ``collective decoherence''
model was introduced (where several
qubits couple identically to the same environment, while undergoing
both dephasing and dissipation), then for general Hamiltonians \cite{Zanardi:97a}. Their
elegant algebraic analysis established the importance of identifying the
dynamical symmetries in the system-environment interaction, and provided the
first general formal condition for decoherence-free (DF) states, that did
not require knowledge of the system-environment coupling strength. In the work of
Zanardi and Rasetti these are referred to as ``error
avoiding codes''. Several papers focusing on
collective dissipation appeared subsequently 
\cite{Zanardi:97b,Duan:98b,Duan:98c}, as well as applications to
encoding information in quantum dots
\cite{Zanardi:98,Zanardi:99aa}. Zanardi \cite{Zanardi:98a}
and independently Lidar, Chuang, and Whaley \cite{Lidar:PRL98} showed that DF states could also be
derived from very general considerations of Markovian master equations. Lidar
 {\it et al}. introduced the term ``decoherence-free subspace'', analyzed their
robustness against perturbations, and pointed out that the absence
of decoherence for DF states can be spoiled by evolution under the system
Hamiltonian, identifying a second major requirement for viable use of the
DF states for either quantum memory or quantum
computation \cite{Lidar:PRL98}. A completely general condition for the
existence of
DF states was subsequently provided in terms of the
Kraus operator sum representation (OSR) by Lidar, Bacon and Whaley \cite{Lidar:PRL99}. All these studies share essentially the same canonical
example of system-environment symmetry: A qubit-permutation
symmetry in the system-environment coupling, that gives rise to collective dephasing,
dissipation, or decoherence.  The other main
example of a symmetry giving rise to large DFSs was provided by Lidar
{\it et al.} in \cite{Lidar:00a}. This is a much weaker form of spatial
symmetry than permutation-invariance, termed ``multiple-qubit
errors'', which we describe in detail in Section \ref{MQE}.

Several papers reported various generalizations of DFSs, e.g., by
extending DFSs to quantum groups \cite{Durdevich:00}, using the rigged
Hilbert space formalism \cite{Qiao:02a}, and deriving DFSs from a
scattering S-matrix approach \cite{Shapiro:02}. However, the next major step forward in generalizing the DFS concept was taken
by Knill, Laflamme, and Viola \cite{Knill:99a}, who introduced the notion of a
``noiseless subsystem''. Whereas previous DFS work had characterized
DF states as singlets (one dimensional irreducible representations) of
the algebra generating the dynamical symmetry in the system-bath
interaction, the work by Knill {\it et al}. showed that higher dimensional
irreducible representations can support DF states as well. An
important consequence was the reduction of the number of qubits needed
to construct a DFS under collective decoherence from four to
three. This was noted independently by De Filippo \cite{DeFilippo:00}
and by Yang and Gea-Banacloche \cite{Yang:01}. The generalization
from subspaces to subsystems has provided a powerful and elegant tool on
which the full theory of universal, fault tolerant quantum computation on
DF states has now been established
\cite{Zanardi:99a,Kempe:00}. It has also provided a basis for unifying
essentially all known decoherence-suppression/avoidance strategies \cite{Zanardi:99d,Zanardi:02}. In
the remainder of this chapter we shall follow our usual convention to refer
to both subsystems and subspaces interchangeably with the acronym DFS, the
distinction being made explicit when necessary. 

Following the initial studies establishing the {\em conditions} for DFSs, Bacon,Lidar, and Whaley made a thorough investigation of the robustness of DF
states to symmetry breaking perturbations \cite{Bacon:99}. These authors also showed
that the passive error correction (``error avoidance'') properties of a DFS
can be combined with the active error correction protocols provided by
quantum error correction by concatenation of a DFS inside a quantum error
correcting code (QECC) \cite{Kitaev:book,Steane:99}, resulting in an encoding capable of protecting
against both collective and independent errors \cite{Lidar:PRL99}. A combined DFS-QECC method was shown to be
necessary in order to be enable universal, fault tolerant quantum
computation in the multiple-qubit errors model
\cite{Lidar:00b}. Interestingly, the DFS for collective decoherence
offers a natural energy barrier against other decoherence processes, a
phenomenon termed ``supercoherence'' by Bacon, Brown and Whaley
\cite{Bacon:01}. Several more recent studies, by Alber {\it et al.} and
Khodjasteh and Lidar, have considered other
hybrid DFS-QECC schemes, focusing in particular on protection against
spontaneous emission \cite{Alber:01,Alber:01a,KhodjastehLidar:02}. DFSs have also been
combined with the quantum Zeno effect by Beige {\it et al.} \cite{Beige:99,Beige:00}. Most
recently, a combination of DFSs and the method of dynamical decoupling
\cite{Viola:98} was shown to offer a complete alternative to QECC \cite{ByrdLidar:01a,LidarWu:02}.

These explicit theoretical demonstrations that, firstly, DF states exist
and can provide stable quantum memory, and second, that fault tolerant universal
computation can be performed on states encoded into a DFS, while initially superficially surprising to many, have
generated several experimental searches for verification of DFSs. The first
experimental verification came with a demonstration by Kwiat {\it et
  al.} of a 2-qubit DFS
protecting photon states against collective dephasing \cite{Kwiat:00}. The
same 2-qubit DFS was subsequently constructed and verified to reduce
decoherence in ion trap experiments by Kielpinski {\it et al.} \cite{Kielpinski:01}. In the latter
experiment an atomic state of one ion was combined with that of a spectator
ion to form a 2-ion DF state that was shown to be protected against
dephasing deriving from long wavelength ambient magnetic field fluctuations.
This experiment is significant in showing the potential of DFS states to
protect fragile quantum state information against decoherence operative in
current experimental schemes. More recently, universal control on the same
2-qubit DFS for collective dephasing has been demonstrated in the context of
liquid state nuclear magnetic resonance by Fortunato {\it et al.} \cite{Fortunato:01}. Nuclear magnetic
resonance has also led to the first experimental demonstration, by
Viola {\it et al.}, of a 3-qubit
DF subsystem, providing immunity against full
collective decoherence deriving from a combination of collective spin flips
and collective dephasing \cite{Viola:01b}. With these first experimental
demonstrations, further experimental efforts towards implementation of DFSs
in active quantum computation, or merely as quantum memory encodings to
transport single qubits from one position to another without incurring
dephasing, seems assured. For example,
the use of a DFS has been identified
as a major component in the construction of a scalable trapped ion
quantum computer \cite{Kielpinski:02,Brown:02,LidarWu:02}.

A common criticism of the theory of DFSs has been that the conditions
required for a DFS to exist are very stringent and the assumptions
underlying the theory may be too unrealistic. It is important to
emphasize that the DFS concept was never meant to provide a full and
independent solution to all decoherence problems. Instead, the central
idea has been {\em to make use of the dynamical symmetries in the
system-environment interaction first} (if they exist), and then to
consider the next level of protection against decoherence. The robustness
properties of DFSs ensure that this is a reasonable approach. In
addition, while the experimental
evidence to date \cite{Kwiat:00,Kielpinski:01,Viola:01b,Fortunato:01,Fortunato:02,Ollerenshaw:02}, is a
reason for cautious optimism, there have also been a number of
theoretical studies showing that the conditions for DFSs may be
{\em created} via the use of the dynamical decoupling method, by
symmetrizing the system-bath interaction \cite{Zanardi:98b,Viola:00a,WuLidar:01b,WuByrdLidar:02}. This holds
true for a wide range of system-bath interaction Hamiltonians. It seems
quite plausible that such active ``environment engineering'' methods
will be necessary for DFSs to become truly comprehensive tools in the
quest to protect fragile quantum information.

In the remainder of this review we will now leave the historical
perspective and provide instead a summary of the theory of DF
subspaces and their generalizations, DF subsystems. We
shall start, in Section \ref{simple} with a simple example of
DFSs in physical systems that is then used as a basis for a rigorous
analysis of decoherence and the conditions for lack of this in an open
quantum system (Section \ref{dec}). We then provide, in Section
\ref{DFS-cond}, a series of
complementary characterizations of what a DFS is, using both exact and
approximate formulations of open systems dynamics. A number of
examples of DFSs in different
physical systems follow in Section \ref{examples}, some of them
new. The later sections of the review deal with the generalization to DF Subsystems
(Section \ref{subsystems}), and the robustness of DFSs
(Section \ref{robustness}). We conclude in Section \ref{conclusions}.

\section{A Simple Example of Decoherence-Free Subspaces: Collective Dephasing
}
\label{simple}

Let us begin by analyzing in detail the operation of the simplest DFS. This
example, first analyzed by Palma \textit{et al.} in \cite{Palma:96} and
generalized by Duan \&\ Guo \cite{Duan:98}, will serve to illustrate what is
meant by a DFS. Suppose that a system of $K$ qubits (two-level
systems) is coupled to a bath in
a symmetric way, and undergoes a dephasing process. Namely, qubit $j$
undergoes the transformation 
\begin{equation}
|0\rangle _{j}\rightarrow |0\rangle _{j}\qquad |1\rangle _{j}\rightarrow
e^{i\phi }|1\rangle _{j},
\end{equation}
which puts a random phase $\phi $ between the basis states $
|0\rangle $ and $|1\rangle $ (eigenstates of $\sigma_z$ with
respective eigenvalues $+1$ and $-1$). This can also be described by the matrix $
R_{z}(\phi )=\mathrm{diag}\left( 1,e^{i\phi }\right) $ acting on the $
\{|0\rangle ,|1\rangle \}$ basis. We assume that the phase has no space ($j$) dependence, i.e., the dephasing process is invariant under qubit
permutations. This symmetry is an example of the more general situation
known as ``collective decoherence'' . Since
the errors can be expressed in terms of the single Pauli spin matrix $\sigma
_{z}$ of the two-level system, we refer to this example as \textquotedblleft
weak collective decoherence'' . The more general situation
when errors involving all three Pauli matrices are present, i.e.,
dissipation and dephasing, is referred to as ``strong
collective decoherence'' . Without encoding a qubit
initially in an arbitrary pure state $|\psi \rangle _{j}=a|0\rangle
_{j}+b|1\rangle _{j}$ will decohere. This can be seen by calculating its
density matrix as an average over all possible values of $\phi $, 
\[
\rho _{j}=\int_{-\infty }^{\infty }R_{z}(\phi )|\psi \rangle _{j}\langle
\psi |R_{z}^{\dagger }(\phi )\,p(\phi )d\phi ,
\]
where $p(\phi )$ is a probability density, and we assume the initial state
of all qubits to be a product state. For a Gaussian distribution, $p(\phi
)=\left( 4\pi \alpha \right) ^{-1/2}\exp (-\phi ^{2}/4\alpha )$, it is
simple to check that 
\[
\rho _{j}=\left( 
\begin{array}{cc}
|a|^{2} & ab^{\ast }e^{-\alpha } \\ 
a^{\ast }be^{-\alpha } & |b|^{2}
\end{array}
\right) .
\]
The decay of the off-diagonal elements in the computational basis is a
signature of decoherence.

Let us now consider what happens in the two-qubit Hilbert space. The four
basis states undergo the transformation 
\begin{eqnarray*}
|0\rangle _{1}\otimes |0\rangle _{2} &\rightarrow &|0\rangle _{1}\otimes
|0\rangle _{2} \\
|0\rangle _{1}\otimes |1\rangle _{2} &\rightarrow &e^{i\phi }|0\rangle
_{1}\otimes |1\rangle _{2} \\
|1\rangle _{1}\otimes |0\rangle _{2} &\rightarrow &e^{i\phi }|1\rangle
_{1}\otimes |0\rangle _{2} \\
|1\rangle _{1}\otimes |1\rangle _{2} &\rightarrow &e^{2i\phi }|1\rangle
_{1}\otimes |1\rangle _{2}\rangle .
\end{eqnarray*}
Observe that the basis states $|0\rangle _{1}\otimes |1\rangle _{2}$ and $
|1\rangle _{1}\otimes |0\rangle _{2}$ acquire the same phase. This suggests
that a simple encoding trick can solve the problem. Let us define encoded
states by $|0_{L}\rangle =|0\rangle _{1}\otimes |1\rangle _{2}\equiv
|01\rangle $ and $|1_{L}\rangle =|10\rangle $. Then the state $|\psi
_{L}\rangle =a|0_{L}\rangle +b|1_{L}\rangle $ evolves under the dephasing
process as 
\[
|\psi _{L}\rangle \rightarrow a|0\rangle _{1}\otimes e^{i\phi }|1\rangle
_{2}+be^{i\phi }|1\rangle _{1}\otimes |0\rangle _{2}=e^{i\phi }|\psi
_{L}\rangle , 
\]
and the overall phase thus acquired is clearly unimportant. This means that
the 2-dimensional subspace $DFS_{2}(0)=\mathrm{Span}\{|01\rangle
,|10\rangle \}$ of the 4-dimensional Hilbert space of two qubits is \emph{decoherence-free}. The subspaces
$DFS_{2}(2)=\mathrm{Span}\{|00\rangle \}$ and $DFS_{2}(-2)=\mathrm{Span}\{|11\rangle \}$ are also (trivially) DF,
since they each acquire a global phase as well, $1$ and $e^{2i\phi }$
respectively. Since the phases acquired by the different subspaces differ,
there is decoherence \emph{between} the subspaces.\footnote{This
  conclusion is actually somewhat too strict: for a non-equilibrium environment, superpositions of different DFSs
can be coherent in the collective dephasing model \cite{Gheorghiu:02}.}

For $K=3$ qubits a similar calculation reveals that the subspaces $
DFS_{3}(1)=\mathrm{Span}\{|001\rangle ,|010\rangle ,|100\rangle \}$ and $
DFS_{3}(-1)=\mathrm{Span}\{|011\rangle ,|101\rangle ,|110\rangle \}$ are DF,
as well the (trivial)\ subspaces $DFS_{3}(3)=\mathrm{Span}\{|000\rangle \}$
and $DFS_{3}(3)=\mathrm{Span}\{|111\rangle \}$.

More generally, let 
\[
\lambda _{K}=\mathrm{number\ of\ }0^{\prime }\mathrm{s\ minus\ the\ number\
of\ }1^{\prime }\mathrm{s} 
\]
in a computational basis state (i.e., a bitstring) over $K$ qubits. Then it
is easy to check that any subspace spanned by states with constant $\lambda
_{K}$ is DF, and can be denoted $DFS_{K}(\lambda _{K})$ in accordance with
the notation above. The dimensions of these subspaces are given by the
binomial coefficients: $d\equiv \dim [DFS_{K}(\lambda _{K})]={K \choose
\lambda _{K}}$ and they each encode $\log _{2}d$ qubits.

The encoding for the ``collective phase
damping''\ model discussed here has been tested experimentally. The first-ever
experimental implementation of DFSs used the $DFS_{2}(0)$ subspace to
protect against artifially induced decoherence in a linear optics setting  
\cite{Kwiat:00}. The same encoding was subsequently used to alleviate the
problem of external fluctuating magnetic fields in an ion trap quantum
computing experiment \cite{Kielpinski:01}, and figures prominently in
theoretical constructions of encoded, universal quantum computation \cite{LidarWu:01,LidarWu:02,ByrdLidar:01a}.

\section{Formal Treatment of Decoherence}
\label{dec}

Let us now present a more formal treatment of decoherence. Consider a closed
quantum system, composed of a system $S$ of interest defined on a Hilbert
space $\mathcal{H}$ (e.g., a quantum computer) and a bath $B$. The full
Hamiltonian is 
\begin{equation}
\mathbf{H}=\mathbf{H}_{S}\otimes \mathbf{I}_{B}+\mathbf{I}_{S}\otimes 
\mathbf{H}_{B}+\mathbf{H}_{I},  \label{eq:H}
\end{equation}
where $\mathbf{H}_{S}$, $\mathbf{H}_{B}$ and $\mathbf{H}_{I}$ are,
respectively, the system, bath and system-bath interaction Hamiltonians, and 
$\mathbf{I}$ is the identity operator. The evolution of the closed system is
given by $\rho _{SB}(t)=\mathbf{U}\rho _{SB}(0)\mathbf{U}^{\dagger }$, where 
$\rho _{SB}$ is the combined system-bath density matrix and the unitary
evolution operator is 
\begin{equation}
\mathbf{U}=\exp (-i\mathbf{H}t)  \label{eq:U}
\end{equation}
(we set $\hbar =1$). Assuming initial decoupling between system and
bath, the evolution of the closed system is given by: $\rho _{SB}(t)=\mathbf{
U}(t)[\rho _{S}(0)\otimes \rho _{B}(0)]\mathbf{U}^{\dagger }(t)$. Without
loss of generality the interaction Hamiltonian can be written as 
\begin{equation}
\mathbf{H}_{I}=\sum_{\alpha }\mathbf{S}_{\alpha }\otimes {\ \mathbf{B}}
_{\alpha },  \label{eq:H_I}
\end{equation}
where $\mathbf{S}_{\alpha }$ and $\mathbf{B}_{\alpha }$ are, respectively,
system and bath operators. It is this coupling between system and bath that
causes decoherence in the system, through entanglement with the bath. To see
this more clearly it is useful to arrive at a description of the system
alone by averaging out the uncontrollable bath degrees of freedom, a
procedure formally implemented by performing a partial trace over the bath: 
\[
\rho (t)={\textrm{Tr}}_{B}[\mathbf{U}(t)\rho (0)\otimes \rho _{B}(0)
\mathbf{U}^{\dagger }(t)].
\]
The reduced density matrix $\rho (t)$ now describes the system alone. By
diagonalizing the initial bath density matrix, $\rho _{B}(0)=\sum_{\nu
}\lambda _{\nu }|\nu \rangle \langle \nu |$, and evaluating the partial
trace in the same basis one finds: 
\begin{eqnarray}
\rho (t) &=&\sum_{\mu }\langle \mu |\mathbf{U}(t)\left( \rho (0)\otimes
\sum_{\nu }\lambda _{\nu }|\nu \rangle \langle \nu |\right) \mathbf{U}
^{\dagger }(t)|\mu \rangle   \nonumber \\
&=&\sum_{a}\mathbf{A}_{a}\,\rho (0)\,\mathbf{A}_{a}^{\dagger },
\label{eq:OSR}
\end{eqnarray}
where the ``\emph{Kraus operators}''\ are
given by:

\begin{equation}
\mathbf{A}_{a}=\sqrt{\lambda _{\nu }}\langle \mu |\mathbf{U}|\nu \rangle
\;;\qquad a=(\mu ,\nu ).  \label{eq:Amunu}
\end{equation}
The expression (\ref{eq:OSR}) is known as the Operator Sum Representation
(OSR) and can be derived from an axiomatic approach to quantum mechanics,
without reference to Hamiltonians \cite{Kraus:83}. Since \textrm{Tr}$\rho
(t)=1$ the Kraus operators satisfy the normalization constraint 
\begin{equation}
\sum_{a}\mathbf{A}_{a}^{\dagger }\mathbf{A}_{a}=\mathbf{I}_{S}.
\label{eq:OSRnorm}
\end{equation}
Because of this constraint it follows that when the sum in Eq.~(\ref{eq:OSR}) includes only one term the dynamics is unitary. \emph{Thus a simple
criterion for decoherence in the OSR is the presence of multiple
independent terms in the sum in Eq.~(\ref{eq:OSR}).}

While the OSR is a formally exact description of the dynamics of the system
density matrix, its utility is somewhat limited because the explicit
calculation of the Kraus operators is equivalent to a full diagonalization
of the high-dimensional Hamiltonian $\mathbf{H}$. Furthermore, the OSR is in
a sense too strict. This is because as a closed-system formulation it
incorporates the possibility that information which is put into the bath
will back-react on the system and cause a recurrence. Such interactions will
always occur in the closed-system formulation (due to the the Hamiltonian
being Hermitian). However, in many practical situations the likelihood of
such an event is extremely small. Thus, for example, an excited atom which
is in a ``cold''\ bath will radiate a photon
and decohere, but the bath will not in turn return the atom back to its
excited state, except via the (extremely long) recurrence time of the
emission process. In these situations a more appropriate way to describe the
evolution of the system is via a quantum dynamical \emph{semigroup master
equation} \cite{Lindblad:76,Alicki:87}. By assuming that (i) the evolution
of the system density matrix is governed by a one-parameter semigroup
(Markovian dynamics), (ii) the evolution is futher ``completely positive'' \cite{Alicki:87}, and (iii) the
system and bath density matrices are initially decoupled, Lindblad \cite{Lindblad:76} has shown that the most general evolution of the system
density matrix $\mathbf{\rho }_{S}(t)$ is governed by the master equation 
\begin{eqnarray}
{\frac{d\mathbf{\rho }(t)}{dt}} &=&-i[\mathbf{\tilde{H}}_{S},\mathbf{\rho }
(t)]+\mathtt{L}_{D}[\mathbf{\rho }(t)]  \nonumber \\
\mathtt{L}_{D}[\mathbf{\rho }(t)] &=&{\frac{1}{2}}\sum_{\alpha ,\beta
=1}^{M}a_{\alpha \beta }\left( [\mathbf{F}_{\alpha },\mathbf{\rho }(t)
\mathbf{F}_{\beta }^{\dagger }]+[\mathbf{F}_{\alpha }\mathbf{\rho }(t),
  \mathbf{F}_{\beta }^{\dagger }]\right)
\label{eq:mastereq}
\end{eqnarray}
where $\mathbf{\tilde{H}}_{S}=\mathbf{H}_{S}+\Delta $ is the system
Hamiltonian $\mathbf{H}_{S}$ including a possible unitary contribution from
the bath $\Delta $ (``Lamb shift'' ), the
operators $\mathbf{F}_{\alpha }$ constitute a basis for the $M$-dimensional
space of all bounded operators acting on $\mathcal{H}$, and $a_{\alpha \beta}$ are the elements of a positive semi-definite Hermitian matrix. The
commutator involving $\mathbf{\tilde{H}}_{S}$ is the ordinary, unitary,
Heisenberg term. All the non-unitary, decohering dynamics is accounted for
by $\mathtt{L}_{D}$, and this is one of the advantages of the Lindblad
equation: unlike the OSR, it clearly separates unitary from decohering
dynamics. For a derivation of the Lindblad equation from the OSR, using a
coarse graining procedure, including an explicit calculation of the
coefficients $a_{\alpha \beta }$ and the Lamb shift $\Delta $, see \cite{Lidar:CP01}. Note that the $\mathbf{F}_{\alpha }$ can often be identified
with the $\mathbf{S}_{\alpha }$ of the interaction Hamiltonian in Eq.~(\ref{eq:H_I}) \cite{Lidar:CP01}.

\section{The DFS Conditions}

\label{DFS-cond}

With the above statement of the conditions for decoherence let us now show
how to formally eliminate decoherence. It is convenient to do so by
reference to the Hamiltonian, OSR, and semigroup formulations. This leads to
a number of essentially equivalent formulations of the conditions for
DF dynamics, whose utility is determined by the approach one
would like to employ to study a specific problem. In addition, it is useful
to give formulations which make contact with the theory of quantum error
correcting codes (QECC). First, let us give a formal definition of a DFS:

\begin{definition}
A system with Hilbert space ${\cal H}$ is said to have a decoherence-free
subspace $\tilde{{\cal H}}$ if the evolution inside $\tilde{{\cal H}}\subset 
{\cal H}$ is purely unitary.
\end{definition}

Note that because of the possibility of a bath-induced Lamb shift [$\Delta $
in Eq.~(\ref{eq:mastereq})] this definition of a DFS does not entirely rule
out adverse effects a bath may have on a system. Also, we are not excluding
unitary errors that may be the result of inaccurate implementation of
quantum logic gates.
Both of these problems, which in practice are inseparable, must be
dealt with by other methods, such as concatenated codes \cite{Lidar:PRL99}.

\subsection{Hamiltonian Formulation}

As remarked above, in terms of the Hamiltonian $\mathbf{H}$ of Eq.~(\ref{eq:H}), decoherence is the result of the entanglement between system and
bath caused by the interaction term $\mathbf{H}_{I}$. In other words, if $
\mathbf{H}_{I}=0$ then system and bath are decoupled and evolve
independently and unitarily under their respective Hamiltonians $\mathbf{H}
_{S}$ and $\mathbf{H}_{B}$. Clearly, then, a sufficient condition for \emph{decoherence free }(DF) dynamics is that $\mathbf{H}_{I}=0$. However, since
one cannot simply switch off the system-bath interaction, in order to
satisfy this condition it is necessary to look for special subspaces of the
full system Hilbert space $\mathcal{H}$. As shown first by Zanardi and
Rasetti \cite{Zanardi:97a}, such a subspace is
found by assuming that there exists a set $\{|\tilde{k}\rangle \}$ of
eigenvectors of the $\mathbf{S}_{\alpha }$'s with the property that:

\begin{equation}
\mathbf{S}_{\alpha }|\tilde{k}\rangle =c_{\alpha }|\tilde{k}\rangle \qquad
\forall \alpha ,|\tilde{k}\rangle .  \label{eq:DFS-cond}
\end{equation}
Note that these eigenvectors are \emph{degenerate}, i.e., the eigenvalue $
c_{\alpha }$ depends only on the index $\alpha $ of the system
operators, but not on the state index $k$.
If $\mathbf{H}_{S}$ leaves the Hilbert subspace $\tilde{\mathcal{H}}=\mathrm{
Span}[\{|\tilde{k}\rangle \}]$ invariant, and if we start within $\tilde{
\mathcal{H}}$, then the evolution of the system will be DF. To show
this we follow the derivation in \cite{Lidar:PRL99}: First expand the initial density matrices of the system and the bath in their
respective bases: $\rho _{S}(0)=\sum_{ij}s_{ij}|\tilde{\imath}\rangle
\langle \tilde{j}| $ and $\rho _{B}(0)=\sum_{\mu \nu }{b}_{\mu \nu }|{\mu }
\rangle \langle \nu | $. Using Eq.~(\ref{eq:DFS-cond}), one can write the
combined operation of the bath and interaction Hamiltonians over $\tilde{
\mathcal{H}}$ as:

\[
\mathbf{I}_{S}\otimes \mathbf{H}_{B}+\mathbf{H}_{I}=\mathbf{I}_{S}\otimes 
\mathbf{H}_{{\textrm{c}}}\equiv \mathbf{I}_{S}\otimes [ \mathbf{H}
_{B}+\sum_{\alpha }a_{\alpha }\mathbf{B}_{\alpha }] . 
\]
This clearly commutes with $\mathbf{H}_S$ over $\tilde{\mathcal{H}}$. Thus
since neither $\mathbf{H}_{S}$ (by our own stipulation) nor the combined
Hamiltonian $\mathbf{H}_{\mathrm{c}}$ takes states out of the subspace:

\begin{equation}
\mathbf{U}[|\tilde{\imath}\rangle \otimes |\mu \rangle ]=\mathbf{U}_{S}|
\tilde{\imath}\rangle \otimes \mathbf{U}_{\mathrm{c}}|\mu \rangle ,
\label{eq:UsUc}
\end{equation}
where $\mathbf{U}_{S}=\exp (-i\mathbf{H}_{S}t)$ and $\mathbf{U}_{{
\textrm{c}}}=\exp (-i\mathbf{H}_{{\textrm{c}}}t)$. Hence it is clear,
given the initially decoupled state of the density matrix, that the
evolution of the closed system will be: $\rho _{SB}(t)=\sum_{ij}s_{ij}
\mathbf{U}_{S}|\tilde{\imath}\rangle \langle \tilde{j}|\mathbf{U}
_{S}^{\dagger }\otimes \sum_{\mu \nu }b_{\mu \nu }\mathbf{U}_{\mathrm{c}
}|\mu \rangle \langle \nu |\mathbf{U}_{\mathrm{c}}^{\dagger }.$ It follows
using simple algebra that after tracing over the bath: $\rho _{S}(t)=\mathrm{
Tr}_{B}[\rho _{SB}(t)]=\mathbf{U}_{S}\rho _{S}(0)\mathbf{U}_{S}^{\dagger }$,
i.e., that the system evolves in a completely unitary fashion on $\tilde{
\mathcal{H}}$: under the condition of Eq.~(\ref{eq:DFS-cond}) the subspace
is DF. As shown in Ref.~ \cite{Zanardi:97a} by performing a short-time
expansion, Eq.~(\ref{eq:UsUc}) is also a \emph{necessary} condition for a
DFS. Let us summarize this:

\begin{theorem}
Let the interaction between a system and a bath be given by the Hamiltonian
of Eq.~(\ref{eq:H}). If ${\bf H}_{S}$ leaves the Hilbert subspace $\tilde{
{\cal H}}={\rm Span}[\{|\tilde{k}\rangle \}]$ invariant, and if we start
within $\tilde{{\cal H}}$, then $\tilde{{\cal H}}$ is a DFS if and only if it
satisfies Eq.~(\ref{eq:DFS-cond}).
\end{theorem}

The condition of Eq.~(\ref{eq:DFS-cond}) is very useful for checking whether
a given interaction supports a DFS: The operators $\mathbf{S}_{\alpha }$
often form a Lie algebra, and the condition (\ref{eq:DFS-cond})\ then
translates into the problem of finding the one-dimensional irreducible
representations (irreps)\ of this Lie algebra, a problem with a textbook
solution \cite{Cornwell:97}. We will consider this in detail through
examples below.

\subsection{Operator-Sum Representation Formulation}

Let $\tilde{\mathcal{H}}$ be an $N$-dimensional DFS. As first observed in 
\cite{Lidar:PRL99}, in this case it follows immediately from Eqs.~(\ref
{eq:UsUc}) and (\ref{eq:Amunu}) that the Kraus operators all have the
following representation (in the basis where the first $N$ states span $
\tilde{\mathcal{H}}$):

\begin{equation}
\mathbf{A}_{a}=\left( 
\begin{array}{cc}
g_{a}\tilde{\mathbf{U}} & \mathbf{0} \\ 
\mathbf{0} & \bar{\mathbf{A}}_{a}
\end{array}
\right) \,;\qquad g_{a}=\sqrt{\nu }\langle \mu |\mathbf{U}_{\mathrm{c}}|\nu
\rangle .  \label{eq:A-block}
\end{equation}
Here $\bar{\mathbf{A}}_{a}$ is an arbitrary matrix that acts on ${\tilde{
\mathcal{H}}}^{\perp }$ ($\mathcal{H}={\tilde{\mathcal{H}}}\oplus {\tilde{
\mathcal{H}}}^{\perp } $) and may cause decoherence there; $\tilde{\mathbf{U}
}$ is restricted to $\tilde{\mathcal{H}}$. This simple condition can be
summarized as follows:

\begin{theorem}
\label{th:Kraus-DFS}A subspace $\tilde{{\cal H}}$ is a DFS if and only if
all Kraus operators have an identical unitary representation upon
restriction to it, up to a multiplicative constant.
\end{theorem}

An explicit calculation will help to illustrate this condition for a DFS.
Consider the set of system states $\{|\tilde{k}\rangle \}$ satisfying: 
\begin{equation}
\mathbf{A}_{a}|\tilde{k}\rangle =g_{a}\tilde{\mathbf{U}}|\tilde{k}\rangle
\qquad \forall a,  \label{eq: DFS-Kraus}
\end{equation}
where $\tilde{\mathbf{U}}$ is an arbitrary, $a$-independent but possibly
time-dependent unitary transformation, and $g_{a}$ a complex constant. Under
this condition, an initially pure state belonging to $\mathrm{Span}[\{|
\tilde{k}\rangle \}]$, 
\[
|\psi _{\mathrm{in}}\rangle =\sum_{k}\gamma _{k}|\tilde{k}\rangle ,
\]
will be DF, since: 
\[
|\phi _{a}\rangle =\mathbf{A}_{a}|\psi _{\mathrm{in}}\rangle =\sum_{k}\gamma
_{k}g_{a}{\tilde{\mathbf{U}}}|\tilde{k}\rangle =g_{a}{\tilde{\mathbf{U}}}
|\psi _{\mathrm{in}}\rangle 
\]
so 
\[
\rho _{\mathrm{out}}=\sum_{a}\mathbf{A}_{a}\tilde{\rho}_{\mathrm{in}}\mathbf{
A}_{a}^{\dagger }=\sum_{a}g_{a}{\tilde{\mathbf{U}}}|\psi _{\mathrm{in}
}\rangle \langle \psi _{\mathrm{in}}|{\tilde{\mathbf{U}}}^{\dagger
}c_{a}^{\ast }={\tilde{\mathbf{U}}}|\psi _{\mathrm{in}}\rangle \langle \psi
_{\mathrm{in}}|{\tilde{\mathbf{U}}}^{\dagger },
\]
where we used the normalization of the Kraus operators [Eq. (\ref{eq:OSRnorm})] to set $\sum_{a}|g_{a}|^{2}=1$. This means that the time-evolved state $
\rho _{\mathrm{out}}$ is pure, and its evolution is governed by ${\tilde{
\mathbf{U}}}$. This argument is easily generalized to an initial mixed state 
$\tilde{\rho}_{\mathrm{in}}=\sum_{kj}\rho _{kj}|\tilde{k}\rangle \langle 
\tilde{j}|$, in which case $\rho _{\mathrm{out}}={\tilde{\mathbf{U}}}\tilde{
\rho}_{\mathrm{in}}{\tilde{\mathbf{U}}}^{\dagger }$. The unitary
transformation ${\tilde{\mathbf{U}}}$ can be exploited in choosing a
driving system Hamiltonian which implements a useful evolution on the DFS.
The calculation above shows that Eq.~(\ref{eq:DFS-cond}) is a sufficient
condition for a DFS. It follows from the results of Refs.~\cite{Zanardi:97a,Nielsen:97} that it is also a necessary condition for a
DFS (under ``generic'' conditions -- to be
explained below).

A useful alternative formulation of the DFS condition using Kraus operators
can be given, that is slightly less general. Consider a group $\mathcal{G}=\{
\mathbf{G}_{n}\}$ and expand the Kraus operators as a linear combination
over the group elements: $\mathbf{A}_{a}=\sum_{n=1}^{N}b_{a,n}\mathbf{G}_{n}$
(i.e., the Kraus operators belong to the group algebra of $\mathcal{G}$).
Then the following theorem gives an alternative DFS characterization \cite{Lidar:00a}:

\begin{theorem}
\label{th:MQE-DFS}If a set of states $\{|\tilde{j}\rangle \}$ belong to a
given \emph{one-dimensional} irrep $\Gamma ^{k}$ of $\mathcal{G}$, then the
DFS condition $\mathbf{A}_{a}|\tilde{j}\rangle =c_{a}|\tilde{j}\rangle $
holds. If no assumptions are made on the bath coefficients $\{b_{a,n}\}$,
then the DFS condition $\mathbf{A}_{a}|\tilde{j}\rangle =c_{a}|\tilde{j}
\rangle $ implies that $|\tilde{j}\rangle $ belongs to a \emph{one-dimensional} irrep $\Gamma ^{k}$ of $\mathcal{G}$.
\end{theorem}

It is not always possible to expand the Kraus operators over a group.
However, when this is possible, the last theorem can be used to find a class
of DFSs under much relaxed symmetry assumptions. This \textquotedblleft
multiple qubit-errors''\ model, introduced in \cite{Lidar:00a} and analyzed for its ability to support universal quantum
computation in \cite{Lidar:00b}, is discussed in more detail in Section \ref{MQE} below.

\subsection{Lindblad-Semigroup Formulation}

Let us now consider the conditions for the existence of a DFS in terms of
the Lindblad semigroup master equation (\ref{eq:mastereq}). The Lindblad equation
nicely separates the unitary and decohering dynamics. It is clear that the
DFS condition should amount to the vanishing of the $\mathtt{L}_{D}[\mathbf{
\rho }]$ term. Let us derive necessary and sufficient conditions for this to
happen, following \cite{Lidar:PRL98}. Let $\{|\tilde{k}\rangle \}_{k=1}^{N}$
be a basis for an $N$-dimensional subspace ${\tilde{\mathcal{H}}}\subseteq 
\mathcal{H}$. In this basis, we may express states as the density matrix 
\begin{equation}
\tilde{\rho}=\sum_{k,j=1}^{N}\rho _{kj}|\tilde{k}\rangle \langle \tilde{j}
|\,.  \label{eq:rho0}
\end{equation}
Consider the action of the Lindblad operators on the basis states: $\mathbf{F
}_{\alpha }|\tilde{k}\rangle =\sum_{j=1}^{N}c_{kj}^{\alpha }|\tilde{j}
\rangle $. Substituting into Eq.~(\ref{eq:mastereq}) we find

\begin{eqnarray}
\mathtt{L}_{D}[\mathbf{\rho }(t)]={\frac{1}{2}}\sum_{\alpha ,\beta
  =1}^{M}a_{\alpha \beta } \times \nonumber \\
\sum_{kj,mn=1}^{N}\rho _{kj}\left( 2c_{jm}^{\beta
\ast }c_{kn}^{\alpha }|\tilde{n}\rangle \langle \tilde{m}
|-c_{mn}^{\beta \ast }c_{kn}^{\alpha }|\tilde{m}\rangle \langle 
\tilde{j}|-c_{jm}^{\beta \ast }c_{nm}^{\alpha }|\tilde{k}\rangle
\langle \tilde{n}|\right) =0.  \label{eq:x1}
\end{eqnarray}
The coefficients $a_{\alpha \beta }$ represent information about the bath,
which we assume is uncontrollable. Hence we must require that each term in
the sum over $\alpha ,\beta $ vanishes separately. Furthermore, we wish to
avoid a dependence on initial conditions, i.e., there should be no
dependence on $\rho _{kj}$. This implies that each of the terms in
parentheses must vanish separately. This can only be achieved if there is
just one projection operator $|\tilde{n}\rangle \langle \tilde{m}|$
in each term. The least restrictive choice leading to this is: $
c_{kn}^{\alpha }=c_{\alpha k}\delta _{kn}$. Eq.~(\ref{eq:x1}) then becomes:

\begin{equation}
\sum_{k,j=1}^{N}\rho _{kj}|\tilde{k}\rangle \langle \tilde{j}|\left(
2c_{j}^{\beta \ast }c_{k}^{\alpha }-c_{k}^{\beta \ast }c_{k}^{\alpha
}-c_{j}^{\beta \ast }c_{j}^{\alpha }\right) =0.  \label{eq:x2}
\end{equation}
Assuming $c_{\alpha k}\neq 0$ then yields: $\frac{c_{\alpha j}}{c_{\alpha k}}
+\frac{c_{\beta k}^{\ast }}{c_{\beta j}^{\ast }}=2$. This has to hold in
particular for $\alpha =\beta $. With $z=c_{\alpha j}{/}c_{\alpha k}$, we
then obtain $z+1/z^{\ast }=2$, which has the unique solution $z=1$. This
implies that $c_{\alpha k}$ must be independent of $k$ and therefore that $
\mathbf{F}_{\alpha }|\tilde{k}\rangle =c_{\alpha }|\tilde{k}\rangle $, $\forall \alpha $. We have thus proved (see also \cite{Zanardi:98a}):

\begin{theorem}
If no special assumptions are made on the coefficient matrix $a_{\alpha
\beta }$ [Eq.~(\ref{eq:mastereq})] and on the initial conditions $\rho _{ij}$
[Eq.~(\ref{eq:rho0})] then a necessary and sufficient condition for a
subspace ${\tilde{{\cal H}}}={\rm Span}[\{|\tilde{k}\rangle
\}_{k=1}^{N}] $ to be decoherence-free is that all basis states $|\tilde{
k}\rangle $ are degenerate eigenstates of all the Lindblad operators $\{{\bf 
F}_{\alpha }\}$ 
\begin{equation}
{\bf F}_{\alpha }|\tilde{k}\rangle =c_{\alpha }|\tilde{k}\rangle
\quad \forall \alpha ,k.  \label{eq:DFD}
\end{equation}
\end{theorem}

The Lindblad operators can always be closed as a Lie algebra $\mathcal{L}$.
We can make the DFS condition somewhat more explicit in the case of \emph{semisimple} Lie algebras, i.e., those which have no Abelian invariant
subalgebra \cite{Cornwell:97}. Using Eq.~(\ref{eq:DFD}) we have $[\mathbf{F}
_{\alpha },\mathbf{F}_{\beta }]|\tilde{k}\rangle =0$. If $\mathcal{L}$
is semisimple then the commutator can be expressed in terms of non-vanishing
structure constants $f_{\alpha ,\beta }^{\gamma }$ of the Lie algebra: $[
\mathbf{F}_{\alpha },\mathbf{F}_{\beta }]=\sum_{\gamma =1}^{M}f_{\alpha
,\beta }^{\gamma }\mathbf{F}_{\gamma }$. We then arrive at the condition on
the structure constants

\begin{equation}
\sum_{\gamma =1}^{M}f_{\alpha ,\beta }^{\gamma }c_{\gamma }=0\quad \forall
\alpha ,\beta .  \label{eq:f}
\end{equation}
The structure constants themselves define the $M$-dimensional adjoint matrix
representation of $\mathcal{L}$ \cite{Cornwell:97}: $\left[ \mathrm{ad}(\hat{
\mathbf{F}}_{\alpha })\right] _{\gamma ,\beta }=f_{\alpha ,\beta }^{\gamma }$. Since the generators of the Lie algebra are linearly independent, so must
be the matrices of the adjoint representation. One can readily show that
this is inconsistent with Eq.~(\ref{eq:f}) unless all $c_{\gamma }=0$. Thus
the DFS condition in the case of semisimple Lie algebras is simply 
\begin{equation}
\mathbf{F}_{\alpha }|\tilde{k}\rangle =0\quad \forall \alpha ,k.
\end{equation}
We will consider examples of this below.

\subsection{Quantum Error Correction Formulation}

Information encoded in a DFS is immune to errors. In QECC information is
encoded into states which can be perturbed by the environment, but can be
recovered. Thus DFSs can be viewed as a special type of ``degenerate''\
QECC, where both the perturbation and recovery are trivial. To formalize
this observation \cite{Lidar:PRL99,Duan:98d}, note that quantum error correction can
be regarded as the theory of reversal of quantum operations on a subspace 
\cite{Nielsen:98}. This subspace, $\mathcal{C}=\mathrm{Span}[\{|i_{L}\rangle
\}]$, is interpreted as a ``code'' (with codewords $\{|i_{L}\rangle \}$)
which can be used to protect part of the system Hilbert space against
decoherence (or ``errors'') caused by the interaction between system and
bath. The errors are represented by the Kraus operators $\{\mathbf{A}_{a}\}$ 
\cite{Knill:97b}. To decode the quantum information after the action of the
bath, one introduces ``recovery'' operators $\{\mathbf{R}_{r}\}$. A QECC is
a subspace $\mathcal{C}$ and a set of recovery operators $\{\mathbf{R}_{r}\}$. Ref.~\cite{Knill:97b} gives two equivalent criteria for the general
condition for QECC. It is possible to correct the errors induced by a given
set of Kraus operators $\{\mathbf{A}_{a}\}$, (i) iff

\begin{equation}
\mathbf{R}_{r}\mathbf{A}_{a}=\left( 
\begin{array}{cc}
\lambda _{ra}\mathbf{I}_{\mathcal{C}} & \mathbf{0} \\ 
\mathbf{0} & \mathbf{B}_{ra}
\end{array}
\right) \qquad \forall r,a ,  \label{eq:RA-block}
\end{equation}
or equivalently, (ii) iff

\begin{equation}
\mathbf{A}_{a}^{\dagger }\mathbf{A}_{b}=\left( 
\begin{array}{cc}
\gamma _{ab}\mathbf{I}_{\mathcal{C}} & \mathbf{0} \\ 
\mathbf{0} & \bar{\mathbf{A}}_{a}^{\dagger }\bar{\mathbf{A}}_{b}
\end{array}
\right) \qquad \forall a,b.  \label{eq:cond2}
\end{equation}
In both conditions the first block acts on $\mathcal{C}$; $\mathbf{B}_{ra}$
and $\bar{\mathbf{A}}_{a}$ are arbitrary matrices acting on $\mathcal{C}
^{\perp }$ ($\mathcal{H}=\mathcal{C}\oplus \mathcal{C}^{\perp }$). Let us
now explore the relation between DFSs and QECCs. First of all, it is
immediate that DFSs are indeed a valid QECC. For, given the (DFS-)
representation of $\mathbf{A}_{a}$ as in Eq.~(\ref{eq:A-block}), it follows
that Eq.~(\ref{eq:cond2}) is satisfied with $\gamma _{ab}=g_{a}^{\ast }g_{b}$. Note, however, that unlike the general QECC case which has a full-rank
matrix $\gamma _{ab}$, in the DFS case this matrix has rank 1 (since the $a^{
\mathrm{th}}$ row equals row 1 upon multiplication by $g_{1}^{\ast
}/g_{a}^{\ast }$). A\ QECC is said to be non-degenerate if it has full rank 
\cite{Gottesman:97,Knill:97b}. A DFS, therefore, is a completely degenerate
QECC \cite{Lidar:PRL99,Duan:98d}. A related characterization of DFSs is in
terms of \emph{distance} \cite{Steane:96a}: a DFS is a QECC with infinite
distance \cite{Knill:99a}, meaning essentially that the errors it is stable
against can have arbitrary strength.

A DFS is an unusual QECC in another way: as we saw above, decoherence does
not affect a perfect DFS \emph{at all}. Since they are based on a
perturbative treatment, other active QECCs (e.g., stabilizer codes \cite{Gottesman:97}) are specifically constructed to improve the fidelity to a
given order in the error rate, which therefore always allows for some
residual decoherence to take place. The absence of decoherence to any order
for a perfect DFS is due to the existence of symmetries in the system-bath
coupling which allow for an \emph{exact} treatment. These symmetries are
ignored by perturbative QECCs either for the sake of generality, or because
they simply do not exist, as in the case of independent couplings. Given a
DFS, the only errors that can take place involve undesired unitary rotations
of codewords (basis states $\{|\tilde{k}\rangle \}$ of $\tilde{\mathcal{H}}$) inside the DFS, due to imprecisions or a bath-induced Lamb shift. Thus,
the complete characterization of DFSs as a QECC is given by the
following \cite{Lidar:PRL99}:

\begin{theorem}
Let ${\cal C}$ be a QECC for error operators $\{{\bf A}_{a}\}$, with
recovery operators $\{{\bf R}_{r}\}$. Then ${\cal C}$ is a DFS iff upon
restriction to ${\cal C}$, ${\bf R}_{r}\propto \tilde{{\bf U}}_{S}^{\dagger
} $ for all $r$.
\end{theorem}

\textit{Proof}. First suppose $\mathcal{C}$ is a DFS. Then by Eqs.~(\ref
{eq:A-block}) and (\ref{eq:RA-block}), 
\[
\mathbf{R}_{r}\left( 
\begin{array}{cc}
g_{a}\tilde{\mathbf{U}}_{S} & \mathbf{0} \\ 
\mathbf{0} & \bar{\mathbf{A}}_{a}
\end{array}
\right) =\left( 
\begin{array}{cc}
\lambda _{ra}\mathbf{I}_{C} & \mathbf{0} \\ 
\mathbf{0} & \mathbf{B}_{ra}
\end{array}
\right) . 
\]
To satisfy this equation, it must be true that 
\[
\mathbf{R}_{r}=\left( 
\begin{array}{cc}
\frac{\lambda _{ra}}{g_{a}}\tilde{\mathbf{U}}_{S}^{\dagger } & \mathbf{C}_{r}
\\ 
\mathbf{D}_{r} & \mathbf{E}_{r}
\end{array}
\right) ,
\]
where $\mathbf{C}_{r}, \mathbf{D}_{r}$ and $\mathbf{E}_{r}$ are
arbitrary. Multiplying, the condition $g_{a} \mathbf{D}_{r} \tilde{\mathbf{U}}_{S} = \mathbf{0}$ implies $\mathbf{D}_{r}=\mathbf{0}$ by unitarity of $\tilde{
\mathbf{U}}_{S}$. Also, since $\bar{\mathbf{A}}_{a}$ is arbitrary,
generically the condition $\mathbf{C}_{r}\bar{\mathbf{A}}_{a}=\mathbf{0}$
implies $\mathbf{C}_{r}=\mathbf{0}$. Thus upon restriction to $\mathcal{C}=
\tilde{\mathcal{H}}$, indeed $\mathbf{R}_{r} \propto \tilde{\mathbf{U}}
^\dagger_{S}$ (by unitarity of $\tilde{\mathbf{U}}_{S}$, $|\lambda
_{ra}/g_{a}|=1$). Now suppose $\mathbf{R}_{r}\propto \tilde{\mathbf{U}}
^\dagger_{S}$. The very same argument applied to $\mathbf{A}_{a}$ in Eq.~(\ref{eq:RA-block}) yields $\mathbf{A}_{a}\propto \tilde{\mathbf{U}}_{S}$
upon restriction to $\mathcal{C}$. Since this is exactly the condition
defining a DFS in Eq.~(\ref{eq:A-block}), the theorem is proved.

Thus in the sense of reversal of quantum operations on a subspace, DFSs are a particularly simple instance of general QECCs,
where upon restriction to the code subspace, all recovery operators are
proportional to the inverse of the system evolution operator. Of
course, underlying this simplicity is an important assumption of
dynamical symmetry.

\subsection{Stabilizer Formulation}

A most useful tool in the theory of QECC is the \emph{stabilizer}
formalism \cite{Gottesman:97}. The stabilizer in QECC is an Abelian subgroup $\mathcal{
S}$ of the Pauli group (the group formed by tensor products of Pauli
matrices on the qubits). It\ allows identification of the errors the code
can correct: The Kraus-operators $\mathbf{A}_{a}$ of
Eq.~(\ref{eq:Amunu}) can be expanded in a basis $\mathbf{E}_{i}$ of ``errors''. Two types of
errors can be dealt with by stabilizer codes: (i) errors $\mathbf{E}
_{i}^{\dagger }\mathbf{E}_{j}$ that anticommute with some $\mathbf{S}\in 
\mathcal{S}$, and (ii) errors that are part of the stabilizer ($\mathbf{E}
_{i}\in \mathcal{S}$). The first class are errors that require active
correction; the second class (ii) are ``degenerate'' errors that do not
affect the code at all. A duality between QECCs and DFSs can be stated
as follows: QECCs were designed primarily to deal with type (i) errors, but can
also be regarded as DFSs for the errors in their stabilizer \cite{Lidar:00a,Lidar:00b}. Conversely, DFSs were designed primarily to deal with
type (ii) errors, but can in principle be used as a QECC against errors that
are type (i) with respect to $\mathcal{S}$. A further use of the stabilizer
is that it allows one to identify sets of fault-tolerant universal gates for
quantum computation \cite{Gottesman:97a}. To this end, and to make a more
explicit connection to QECCs, it is useful to recast the DFS condition Eq.~(\ref{eq:DFS-cond}) into the stabilizer formalism. By analogy to QECC, we
define the \emph{DFS stabilizer} $\mathcal{S}$ as a set of operators $
\mathbf{D}_{\beta }$ which act as identity on the DFS states: 
\begin{equation}
\mathbf{D}_{\beta }|\psi \rangle =|\psi \rangle \quad \forall \mathbf{D}
_{\beta }\in \mathcal{S}\quad \mathrm{iff}\,\quad |\psi \rangle \in \mathrm{
DFS}.  \label{eq:superdef}
\end{equation}
Here $\beta $ can be a discrete or continuous index; $\mathcal{S}$ can form
a finite set or group. This $\mathcal{S}$ is therefore a generalization of
the QECC stabilizers. While some DFSs can also be specified by a stabilizer
in the Pauli-group \cite{Lidar:00a}, many DFSs are specified by non-Abelian
groups, and hence are \emph{nonadditive} codes \cite{Rains:97}.

Consider now the following continuous index stabilizer: 
\begin{equation}
\mathbf{D}(v_{0},v_{1},...,v_{A})=\mathbf{D}(\vec{v})=\exp \left[
\sum_{\alpha =1}^{A}\left( c_{\alpha }\mathbf{I}-\mathbf{S}_{\alpha }\right)
v_{\alpha }\right] .  \label{eq:DFSstab}
\end{equation}
Clearly, the DFS condition [Eq.~(\ref{eq:DFS-cond})] implies that $\mathbf{D}
(\vec{v})|\psi \rangle =|\psi \rangle $. Conversely, if $\mathbf{D}(\vec{v}
)|\psi \rangle =|\psi \rangle $ for all $\vec{v}$, then in particular it
must hold that for each $\alpha $, $\exp \left[ \left( c_{\alpha }\mathbf{I}-
\mathbf{S}_{\alpha }\right) v_{\alpha }\right] |\psi \rangle =|\psi \rangle $. Recalling that $\phi (\mathbf{A)=}\exp \mathbf{[A}]$ is a one-to-one
continuous mapping of a small neighbourhood of the zero matrix $\mathbf{0}$
onto a small neighbourhood of the identity matrix $\mathbf{I}$, it follows
that there must be a sufficiently small $v_{\alpha }$ such that $\left(
c_{\alpha }\mathbf{I}-\mathbf{S}_{\alpha }\right) |\psi \rangle =0$.
Therefore the DFS condition (\ref{eq:DFS-cond}) holds iff $\mathbf{D}(\vec{v}
)|\psi \rangle =|\psi \rangle $ for all $\vec{v}$.

\subsection{Relative Merits of the Various Formulations}

We have given five different formulations of the conditions for a DFS. The
OSR formulation is the most general. The QECC and stabilizer formulations
are particularly useful for constructing logical operations that correspond
to quantum computation inside a DFS. The unifying theme is \emph{dynamical
symmetry}. A DFS exists if and only if there is a symmetry in the
system-bath coupling. This is made explicit in the Hamiltonian and semigroup
formulations, where the error operators $\mathbf{S}_{\alpha }$ and $\mathbf{F
}_{\alpha }$ span a Lie algebra. The DFS condition then reduces to a search
for the one-dimensional irreps of this algebra.

One may
wonder whether the Hamiltonian and semigroup formulations are equivalent. There
is in fact an important difference between the $\mathbf{S}_{\alpha }$'s and
the $\mathbf{F}_{\alpha }$'s which makes the two decoherence-freeness
conditions different. In the Hamiltonian formulation of DFSs, the
Hamiltonian is Hermitian. Thus the expansion for the interaction Hamiltonian
Eq.~(\ref{eq:H_I}) can always be written such that the $\mathbf{S}_{\alpha }$
are also Hermitian. On the other hand, the $\mathbf{F}_{\alpha }$'s in the
master equation, Eq.~(\ref{eq:mastereq}), need not be Hermitian. Because of
this difference, Eq.~(\ref{eq:DFD}) allows for a broader range of subspaces
than Eq.~(\ref{eq:DFS-cond}). For example, consider the situation where
there are only two nonzero terms in a master equation for a two-level
system, corresponding to $\mathbf{F}_{1}=\sigma _{-}$ and $\mathbf{F}
_{2}=\sigma _{z}$ where $\sigma _{-}=|0\rangle \langle 1|$ and $\sigma
_{z}=|0\rangle \langle 0|-|1\rangle \langle 1|$ (e.g., cooling with phase
damping). In this case there is a DFS corresponding to the single state $
|0\rangle $. In the Hamiltonian formulation, inclusion of $\mathbf{S}
_{1}=\sigma _{-}$ in the interaction Hamiltonian expansion Eq.~(\ref{eq:H_I}) would necessitate a second term in the Hamiltonian with $\mathbf{S}
_{2}=\sigma _{-}^{\dagger }=\sigma _{+}$, along with the $\mathbf{S}
_{z}=\sigma _{z}$ as above. For this set of operators, however, Eq.~(\ref{eq:DFS-cond}) allows for no DFS.

\section{Further Examples of Decoherence-Free Subspaces}
\label{examples}

\subsection{Electromagnetically Induced Transparency\protect\footnote{
We are indebted to Dr. J.C. Garrison for first suggesting the 3-level
version of this example in Dec. 1998.}}

\label{eit}

\begin{figure}
\centering
\includegraphics[height=16cm,angle=270]{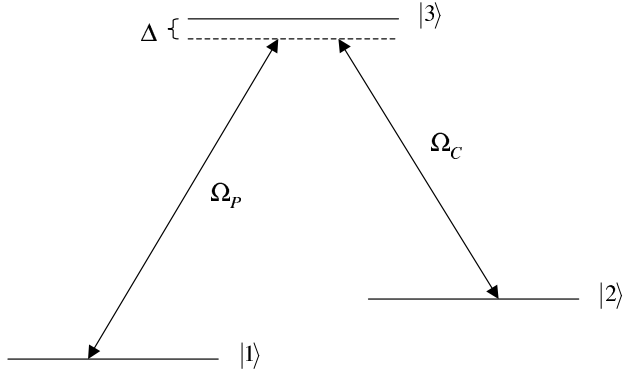}
\vspace{-7cm}
\caption{Level diagram for EIT. A DFS can be formed from levels
  $|1\rangle$ and $|2\rangle$.}
\label{fig:EIT}
\end{figure}

A simple, previously unpublished example of a situation in which a DFS can
arise is the phenomenon of electromagnetically induced transparency
(EIT) \cite{Harris:97}. The aspect of EIT that is of interest to us
here is the formation of a state that is immune to spontaneous
emission. A concrete physical model is provided by a tenuous vapor
of 3-level atoms in the ${\rm \Lambda} $ configuration, e.g., Strontium, contained
in an optical resonator. It is convenient to choose a ring-resonator
geometry in which there are travelling, as opposed to standing, modes. These
modes will be strongly confined, while modes propagating in directions
transverse to the ring path will be essentially unconfined. The atomic
levels are denoted by $\left\vert {1}\right\rangle ,\left\vert {2}
\right\rangle ,\left\vert {3}\right\rangle $, with energies ${\varepsilon }
_{1}\leq \ {\varepsilon }_{2}<\ {\varepsilon }_{3}$. The transitions $(3\
\leftrightarrow 1)$ and $(3\ \leftrightarrow 2)$ are (electric)-dipole
allowed, so $(1\ \leftrightarrow 2)$ is strongly forbidden (by parity:
levels $1$ and $2$ must have equal parity, opposite to that of level $3$).
We assume that collisional broadening can be neglected. In this limit, the
only dissipative (decohering) mechanism is spontaneous emission into the
transverse modes. The atoms are exposed to two laser fields (treated
classically). The \emph{coupling}-laser, with Rabi frequency ${{\rm
    \Omega} }_{C}$, is slightly detuned from the $(3 \leftrightarrow 2)$ transition, and the 
\emph{probe}-laser with Rabi frequency ${{\rm \Omega} }_{P}$ is slightly detuned from the $(3 \leftrightarrow 1)$ transition. See Fig.~\ref{fig:EIT}. However, rather than restricting ourselves to the 3-level
atoms case, suppose that we have an atom with $N+1$ levels, denoted by $
\left\vert {1}\right\rangle ,\left\vert {2}\right\rangle ,...,\left\vert {N+1
}\right\rangle $, with energies ${\varepsilon }_{1}\leq \ {\varepsilon }
_{2}<\ ...<{\varepsilon }_{N+1}$, such that the transitions $
(N+1\leftrightarrow i)$, with $i=1..N$ are allowed, but
$(i\leftrightarrow j)$, with $i,j \in \{1,...,N\}$, are strongly forbidden. This is an artificial model whose
purpose is to illustrate the appearance of a large DFS. The standard EIT
case is recovered by letting $N=2$.

The effective system Hamiltonian is found by (a) transforming to the
interaction picture, (b) imposing the rotating-wave approximation, (c) a
further transformation to remove the remaining explicit time dependence from
the Hamiltonian. In this rotating-wave picture
\cite{Shore:book,Berman:94} the result
in the ordered basis $\{\left\vert {i}\right\rangle \}_{i=1}^{N+1}$ is: 
\begin{eqnarray}
\mathbf{H}_{S} &=& \left( 
\begin{array}{ccccc}
\Delta +\delta _{1} & 0 & \cdots  & 0 & -{{{\rm \Omega} }_{1}/2} \\ 
0 & \Delta +\delta _{2} &  &  & -{{{\rm \Omega} }_{2}/2} \\ 
\vdots  &  & \ddots  &  & \vdots  \\ 
0 &  &  & \Delta  & -{{{\rm \Omega} }_{N}/2} \\ 
-{{{\rm \Omega} }_{1}^{{}}/2} & -{{{\rm \Omega} }_{2}^{{}}/2} & \cdots  & -{{{\rm \Omega} }
_{N}^{{}}/2} & 0
\end{array}
\right) \nonumber \\
&=& \sum_{i=1}^{N}(\Delta +\delta _{i})\mathbf{T}_{ii}-\frac{1}{2}
\sum_{i=1}^{N}{\rm \Omega} _{i}\mathbf{T}_{i,N+1}+{{{\rm \Omega} }_{i}}\mathbf{T}_{N+1,i}
\end{eqnarray}
where the \emph{transition operators} $\{\mathbf{T}_{ab}\}$ are
defined by
\[
\mathbf{T}_{ab}=\left\vert {a}\right\rangle \left\langle {b}\right\vert
,\quad (a,b\in \left\{ {1,...,N+1}\right\} ),
\]
${\rm \Omega} _{i}$ is the Rabi frequency of a
laser field coupling levels $|i\rangle $ and $|N+1\rangle $, $\Delta ={\rm \Omega}
_{i}-\omega _{N+1,i}$ is the coupling laser detuning (assumed independent of 
$i$), and $\delta _{i}={\rm \Omega} _{i+1}-{\rm \Omega} _{i}-\omega _{i+1,i}$ is the
Raman detuning. In standard EIT ${{{\rm \Omega} }_{1}={\rm \Omega} }_{P}$ and ${{\rm \Omega} }
_{2}={{\rm \Omega} }_{C}$.

Decoherence is described in this model by spontaneous emission from the
excited state $|N+1\rangle $ to any of the lower lying states $|i\rangle $,
where $i=1..N$. It is this decoherence process that we wish to avoid. Thus the relevant Lindblad operators $\{\mathbf{F}_{\alpha}\}$ are $\{\mathbf{T}_{i,N+1}\}$ and the $a_{\alpha \beta }$ matrix of Eq.~(\ref{eq:mastereq}) is diagonal:\ $a_{ii}=A_{N+1,i}$, where $A_{N+1,i}$ is
the Einstein $A$ -coefficient for the spontaneous transition from $
|N+1\rangle $ to $|i\rangle $. The DFS condition, Eq.~(\ref{eq:DFD}), now
tells us that if ${\left| {\psi }\right\rangle \in }\tilde{\mathcal{H}}$,
then 
\begin{equation}
\mathbf{T}_{i,N+1}{\left| {\psi }\right\rangle =c}_{i}{\left| {\psi }
\right\rangle .}  \label{eq:DFS-multi}
\end{equation}
However, note that while the operators $\{\mathbf{T}_{ij}\}$ ($i,j=1..N+1$)
generate the full semisimple Lie algebra $su\left( {N+1}\right) $, the $N$
operators $\{\mathbf{T}_{i,N+1}\}$ appearing in the decoherence term of the
Lindblad equation, commute. They span the Abelian Cartan subalgebra
[generally $su\left( {N+1}\right) $ has an $N$-dimensional Cartan
subalgebra, defined as the maximal set of commuting generators] \cite
{Cornwell:97}. This algebra is not semisimple (since it is an Abelian
invariant subalgebra of itself) and the DFS theorem does not tell us what
the $\{{c}_{i}\}$ are. However, a direct way to see this is to calculate the
scalar product of (\ref{eq:DFS-multi}) with itself ($i\neq j$ ): 
\begin{equation}
c_{j}^{\ast }c_{i}=\langle \psi |\mathbf{T}_{j,N+1}^{\dagger }\mathbf{T}
_{i,N+1}{\left| {\psi }\right\rangle =}\langle \psi |N+1\rangle \langle
j|i\rangle \langle N+1|\psi \rangle =0,
\end{equation}
so that either $c_{i}=0$, or $c_{j}=0$, or both. Suppose $c_{j}=0$; then by
Eq.~(\ref{eq:DFS-multi}) the most general form $|\psi \rangle $ can have is: 
\begin{equation}
|\psi \rangle =\sum_{j=1}^{N}a_{j}|j\rangle ,  \label{eq:psi-DFS}
\end{equation}
i.e., $|\psi \rangle $ cannot have a component $|N+1\rangle $:\ $\langle
N+1|\psi \rangle =0$. Thus $c_{i}=0$ for all $i$. This result for the
allowed DFS states is intuitively obvious:\ since spontaneous emission takes
place only from level $N+1$, the DF states
are those which do not contain an $|N+1\rangle $ component.

Let us now take into account the requirement that the DFS be invariant under
the system Hamiltonian. Invariance implies $\mathbf{T}_{k,N+1}(\mathbf{H}
_{S}|\psi \rangle )=0$, so that by Eq.~(\ref{eq:DFS-multi}), a necessary
condition is 
\begin{equation}
\lbrack \mathbf{T}_{k,N+1},\mathbf{H}_{S}]|\psi \rangle =0.  \label{eq:comm}
\end{equation}
Noting that $[\mathbf{T}_{k,N+1},\mathbf{T}_{ij}]=\delta _{i,N+1}|k\rangle
\langle j|-\delta _{jk}|i\rangle \langle N+1|$, we can evaluate the
commutator in Eq.~(\ref{eq:comm}) as follows: 
\begin{eqnarray*}
\lbrack \mathbf{T}_{k,N+1},\mathbf{H}_{S}] &=&\sum_{i=1}^{N}(\Delta +\delta
  _{i})[\mathbf{T}_{k,N+1},\mathbf{T}_{ii}] \nonumber \\
&-& \frac{1}{2}\sum_{i=1}^{N}{\rm \Omega}
_{i}[\mathbf{T}_{k,N+1},\mathbf{T}_{i,N+1}]+{{{\rm \Omega} }_{i}[}\mathbf{T}
_{k,N+1},\mathbf{T}_{N+1,i}]  \nonumber \\
  &=&-\sum_{i=1}^{N}(\Delta +\delta _{i})\delta _{ik}|i\rangle \langle
  N+1|\nonumber \\
&+&
\frac{1}{2}\sum_{i=1}^{N}{\rm \Omega} _{i}\delta _{N+1,k}|i\rangle \langle N+1|+{{
{\rm \Omega} }_{i}}\left( \delta _{ik}|N+1\rangle \langle N+1|-|k\rangle \langle
i|\right) .
\end{eqnarray*}
Using $\langle N+1|\psi \rangle =0$, we obtain: 
\begin{equation}
\lbrack \mathbf{T}_{k,N+1},\mathbf{H}_{S}]|\psi \rangle =-\frac{1}{2}\left[
\sum_{i=1}^{N}{{{\rm \Omega} }_{i}}\langle i|\psi \rangle \right] |k\rangle =0.
\end{equation}
This leads to a \emph{generalized EIT condition}: 
\begin{equation}
\sum_{i=1}^{N}{{{\rm \Omega} }_{i}}\langle \psi |i\rangle =0,  \label{eq:psi-EIT}
\end{equation}
which expresses a destructive interference between all $N$ lower lying
levels. The corresponding superposition suffers no spontaneous
emission, and is preserved under its system Hamiltonian. The standard EIT condition, for $N=2$, is 
\begin{equation}
{{\rm \Omega} }_{P}\langle {\psi }\left\vert {1}\right\rangle +{{\rm \Omega} }
_{C}\langle {\psi }\left\vert {2}\right\rangle =0  \label{eq:EIT}
\end{equation}
also known as the \emph{dark-state}-condition \cite{Harris:97}.

The $N-2$
dimensional DFS defined by Eqs.~(\ref{eq:psi-DFS}),(\ref{eq:psi-EIT}) may be
useful for quantum memory applications. However, it is not clear how to make
use of this DFS for quantum \emph{computing} applications, since all states
in this DFS are degenerate eigenvectors of $\mathbf{H}_{S}$, so that the
evolution inside the DFS is trivial, i.e., a global
phase.\footnote{This issue has been studied in detail and
further generalized by Bacon to other multi-level atom spectra \cite{Bacon:thesis}.} This further
implies that the Raman detunings\ $\delta _{j}$ must all vanish in order for
the DFS to be preserved under $\mathbf{H}_{S}$. Finally, the symmetry that
characterizes this EIT-DFS is the coupling of all spontaneous emission
operators to a common excited state. A DFS would not exist if all levels
were radiatively coupled to each other.

\subsection{Spin Boson Model with Strong Collective Decoherence}
\label{spin-boson}

A beautiful example of a DFS with applications to quantum computing comes
from the spin-boson model \cite{Leggett:87}. This DFS example was proposed
by Zanardi and Rasetti in the influential paper \cite{Zanardi:97c}. Consider 
$N$ spins interacting with a bosonic field via the Hamiltonian
\begin{equation}
\mathbf{H}_{I}=\sum_{i=1}^{N}\sum_{k}\left( g_{i,k}^{+}\sigma
_{i}^{+}\otimes b_{k}+g_{i,k}^{-}\sigma _{i}^{-}\otimes b_{k}^{\dagger
}+g_{i,k}^{z}\sigma _{i}^{z}\otimes ( b_{k}+b_{k}^{\dagger })\right) .
\label{eq:sb_coll}
\end{equation}
Here $\{\sigma _{i}^{+},\sigma _{i}^{-},\sigma _{i}^{z}\}$ are Pauli
operators acting on the $i^{\mathrm{th}}$ spin, $b_{k}$ ($b_{k}^{\dagger }$)
is an annihilation (creation) operator for the $k^{\mathrm{th}}$ bosonic
mode, and $g_{i,k}^{\alpha }$ are coupling constants. The Hamiltonian $
\mathbf{H}_{I}$ describes a rather general interaction between a system of
qubits (the spins) and a bath of bosons, exchanging energy through the $
\sigma _{i}^{+}\otimes b_{k}$ and $\sigma _{i}^{-}\otimes b_{k}^{\dagger }$
terms, and changing phase through the $\sigma _{i}^{z}\otimes (b_{k}+b_{k}^{\dagger })$ term. As it stands $\mathbf{H}_{I}$ does not
support a DFS:\ there are $3N$ $\mathbf{S}_{\alpha }$ operators \ when
comparing to Eq.~(\ref{eq:H_I}), i.e., the $N$ triples of \emph{local} $sl(2)$ algebras $\{\sigma _{i}^{+},\sigma _{i}^{-},\sigma _{i}^{z}\}$; each such
algebra acts on a single qubit, and therefore has a two-dimensional irrep.
The overall action of the total Lie algebra $\bigoplus_{i=1}^{N}sl_{i}(2)$
is represented by the irreducible $N$-fold tensor product of all local
two-dimensional irreps. This implies that there are no one-dimensional
irreps as required by Eq.~(\ref{eq:DFS-cond}), and hence no DFS.

The situation changes dramatically when a \emph{permutation symmetry} is
imposed on the system-bath interaction, in the form 
\begin{equation}
g_{i,k}^{\alpha }\equiv g_{k}^{\alpha }.  \label{eq:gi}
\end{equation}
This ``collective decoherence'' situation is relevant in a number of
solid-state quantum computing proposals, in particular those where
decoherence due to interaction with a cold phonon bath is dominant. At low
temperatures only long-wavelength phonons survive (since there is an energy
gap for excitation of high-energy/short-wavelength phonons), and the model
of collective decoherence becomes relevant provided the qubit spacing is
small compared to the phonon wavelength. Another prototypical situation is
the Dicke model in quantum optics \cite{Dicke:54}, discussed in more detail
below.

Given the collective decoherence assumption, one can define three collective
spin operators $\mathbf{S}_{\alpha }=\sum_{i=1}^{N}\sigma _{i}^{\alpha }$ so
that the Hamiltonian becomes 
\[
\mathbf{H}_{I}=\sum_{\alpha =+,-,z}\mathbf{S}_{\alpha }\otimes \mathbf{B}
_{\alpha }, 
\]
where $\mathbf{B}_{+}=\sum_{k}g_{k}^{+}\sigma _{i}^{+}\otimes b_{k}$, $
\mathbf{B}_{-}=\mathbf{B}_{+}^{\dagger }$ and $\mathbf{B}_{z}=
\sum_{k}g_{i,k}^{z}\sigma _{i}^{z}\otimes ( b_{k}+b_{k}^{\dagger}) $. In fact the specific form of the bath operators $\mathbf{B}
_{\alpha }$ is irrelevant. The important point is that the system operators $
\mathbf{S}_{\alpha }$ now form a \emph{global} $sl(2)$ angular momentum
algebra, i.e., $[\mathbf{S}_{+},\mathbf{S}_{-}]=\mathbf{S}_{z}$ and $[
\mathbf{S}_{z},\mathbf{S}_{\pm }]=2\mathbf{S}_{\pm }$, with a highly \emph{reducible} $2^{N}\times 2^{N}$ representation formed by its action on all $N$
qubits at once. We refer to the case of a global $sl(2)$ algebra as ``strong
collective decoherence''; the case of collective phase damping discussed in
Section~\ref{simple} corresponds to having just a single global angular
momentum operator ($\mathbf{S}_{z}$) and is referred to as ``weak collective
decoherence'' \cite{Kempe:00}. Since $sl(2)$ is semisimple Eq.~(\ref{eq:DFS-cond}) now tells us that the DFS is made up of those states $|\psi
\rangle $ satisfying 
\[
\mathbf{S}_{\alpha }|\psi \rangle =0\qquad \forall \alpha . 
\]
These are the states with vanishing total (fictitious) angular momentum $J$,
i.e., the ``singlets'' of $sl(2)$. Their explicit form is well known for the
case $N=2$, in which case there is only one singlet: 
\[
|\psi \rangle _{N=2}\equiv |s\rangle _{12}=\frac{1}{\sqrt{2}}\left(
|01\rangle -|10\rangle \right) =|J=0,m_{J}=0\rangle 
\]
where the notation $|s\rangle _{12}$ should be read as ``singlet state of
qubits 1 and 2''. This is also known as a Bell state or EPR pair \cite{Nielsen:book}. The final form uses the total angular momentum $J$ and its
projection $m_{J}$. It is easy to check that the state 
\[
|\psi \rangle _{N}=\bigotimes_{m=1}^{N/2}|s\rangle _{2m-1,2m} 
\]
is in the $N$-qubit DFS (since adding pairwise $|J=0,m_{J}=0\rangle $ states
again produced a state with total $J=0$), and that there are no DFS states
for $N$ odd (since such a state always has half-integer total $J$). The
method of Young tableaux can be used to derive a dimensionality formula \cite
{Zanardi:97c,Lidar:PRA00Exchange}:
\begin{equation}
\dim [\mathrm{DFS}(N)]=\frac{N!}{(N/2+1)!(N/2)!}  \label{eq:dim_cd}
\end{equation}
is the number of singlet states for $N$ qubits. An interesting consequence
is that the \emph{encoding efficiency} $\epsilon $ of the DFS, defined as
the number of logical qubits $\log _{2}\dim [\mathrm{DFS}(N)]$ per number of
\ physical qubits $N$, tends asymptotically to unity \cite{Zanardi:97c}: 
\[
\epsilon \stackrel{N \rightarrow \infty }{\longrightarrow }1-\frac{3}{2}\frac{
\log _{2}N}{N}. 
\]
There are several methods for calculating the singlet states for arbitrary $N$, e.g., by solving the problem using linear algebra, using group
representation theory, or by using angular momentum addition rules. To
illustrate the latter, consider the case of $N=4$. The dimensionality
formula yields $\dim [\mathrm{DFS}(4)]=2$, meaning that this DFS \emph{encodes one logical qubit}. One of the states is $|0_{L}\rangle \equiv |\psi
\rangle _{4}=|s\rangle _{12}\otimes |s\rangle _{34}$. The second,
orthogonal, state must be a combination of the triplet states $|t_{-}\rangle
\equiv |00\rangle =|J=1,m_{J}=-1\rangle $, $|t_{+}\rangle \equiv |11\rangle
=|J=1,m_{J}=1\rangle $, $|t_{0}\rangle \equiv \frac{1}{\sqrt{2}}\left(
|01\rangle +|10\rangle \right) =|J=1,m_{J}=0\rangle $. Indeed, the
combination $|1_{L}\rangle =\frac{1}{\sqrt{3}}\left( |t_{-}\rangle
_{12}\otimes |t_{+}\rangle _{34}-|t_{0}\rangle _{12}\otimes |t_{0}\rangle
_{34}+|t_{+}\rangle _{12}\otimes |t_{-}\rangle _{34}\right) $ clearly has
total $J=0$, and has the correct Clebsch-Gordan coefficients. In this manner
one can construct total $J=0$ states for progressively higher $N$.

The encoding which we just
discussed, of logical qubits into blocks of four spins each, was used to
provide the first constructive procedure for universal, fault tolerant quantum
computation using DFSs. As shown by Bacon {\it et al.}
\cite{Bacon:99a}, universality can be achieved by switching on/off only Heisenberg
exchange interactions between pairs of spins. This established for the
first time that the Heisenberg interaction is all by itself universal
for quantum computation. This observation is very important for a
class of solid-state quantum computing proposals, such as electron
spins in quantum dots \cite{Loss:98} and donor atom spins in Si
\cite{Kane:98}, where it is technically preferable to have to control
only a single interaction. The universality of the Heisenberg
interaction was further proved for encodings into any number
$N\geq 3$ of spins in \cite{Kempe:00}. Explicit pulse sequences for
the $N=3$ case were derived in \cite{DiVincenzo:00a}. Furthermore, the
collective decoherence conditions can be {\em created} from arbitrary linear
system-bath interaction Hamiltonians, using strong and fast pulses of
the Heisenberg interaction \cite{WuLidar:01b}, and leakage from a DFS
can be detected using a Heisenberg-only circuit \cite{Kempe:01}, or
prevented using Heisenberg-only pulses \cite{WuByrdLidar:02}.

\subsection{DFS and Dicke subradiance}

A different physical situation in which DFSs arise as a consequence of a
collective coupling to the environment appears in the context of the
Jaynes-Cummings Hamiltonian for $N$ identical two-level atoms coupled to a
single mode radiation field. The Hamiltonian is very similar to that of Eq.~(\ref{eq:sb_coll}), namely
\begin{equation}
\mathbf{H}_{I}=\sum_{i=1}^{N}\left( g_{i}^{+}\sigma _{i}^{+}\otimes
b+g_{i}^{-}\sigma _{i}^{-}\otimes b^{\dagger }\right)   \label{eq:dicke_ham}
\end{equation}
When the atoms are closer together than the wavelength of the radiation
field, then the coupling parameter $g_{i}$ becomes independent of the atom
index $i$, and one can write Eq.~(\ref{eq:dicke_ham}) in terms of the
collective spin operators $\mathbf{S}_{+}=\sum_{i=1}^{N}\sigma _{i}^{+}$ and 
$\mathbf{S}_{-}=\sum_{i=1}^{N}\sigma _{i}^{-}$. The DFS condition may be
applied directly in the Hamiltonian formulation. Together, these two
conditions on the $N$-particle atomic states are equivalent to the single
condition $\mathbf{S}|\psi \rangle =0$ that we derived above for the
collective decoherence of the spin-boson model in the strong coupling
situation (\textit{i.e.}, $\alpha =1,2,$ and $3$). With the atomic states
alone, we therefore similarly arrive at a set of DF states that
are identified with the $N$-atom singlet states. This model was also treated
extensively in \cite{Duan:98b,Duan:98c,Zanardi:98a,Zanardi:97b}. 

The Jaynes-Cummings Hamiltonian is important also in the context of the
Dicke states of quantum optics \cite{Dicke:54}, where the motivation is very
different from ours. The Dicke states are collective atomic states that are
useful in consideration of the treatment of radiation by a collection of $N$
two-level atoms. Let $m\equiv \frac{1}{2}[($number of atoms in the excited
state$)-($number in the ground state$)]$. The Dicke states are defined as
eigenstates of the collective angular momentum, itself defined by $\mathbf{S}^{2}=\mathbf{
S}_{x}^{2}+\mathbf{S}_{y}^{2}+\mathbf{S}_{z}^{2}$, with eigenvalues $|m|\leq
J\leq N/2$, where the effective total angular momentum $J$ is referred to by
Dicke as the ``cooperation number''\ \cite{Dicke:54,Mandel:95}. Dicke showed that the rate of photon emission is given
by $\Gamma =A(J+m)(J-m+1)$, where $A$ is the single-atom Einstein
coefficient. Dicke's interest was primarily in the \textquotedblleft
superradiant''\ states with $J=N/2$ and $m=0$. In this case
the rate of emission is proportional to $N^{2}$, which is unusual since it
implies that the radiation from a partially excited atomic system ($m=0$)
can be much higher than that from a fully excited system ($m=N/2$). It also
follows that there are ``anti-superradiant'', or ``subradiant''\ states, with $\Gamma =0$, corresponding to the situations $m=-J$ and $J=m=0$ \cite{Dicke:54,Mandel:95}.\footnote{
It is important to distinguish subradiant states from \textquotedblleft
dark''\ states, which are also well known in quantum optics 
\cite[p.824]{Shore:book}. A dark state is a special dressed superposition
state of the excited states of a \emph{single} atom that is not dipole
coupled to the ground state of the atom even if the undressed excited states
themselves are. A dark state is usually defined in terms of stationary light
fields coupled to separate transitions with a common ground state. It is
created if the detunings $\delta _{i}=\epsilon _{i}-\omega _{i}$ of the two
light fields with frequencies $\omega _{i}$ from their respective transition
frequencies $\epsilon _{i}$ in a V or ${\rm \Lambda} $ configuration are equal.
Note the similarity to the ``dark-state
condition''\ of EIT, Eq.~(\ref{eq:EIT}), where equal
detunings were also assumed, albeit at the level of Rabi frequencies.} Evidence of this
behavior has been obtained experimentally for trapped ions \cite{DeVoe:96}.
The case $J=m=0$ coincides with that of a DFS spanned by singlet states.
However, one should not conclude that DFSs are nothing but Dicke's
subradiant states. First of all, there is a major conceptual difference,
since in DFS theory we consider the radiation field as an external
environment, while in Dicke's analysis the notion of an environment did not
play a central role. In fact the explanation of subradiance is
straightforward: the atomic dipoles are oppositely phased in pairs, so that
the field radiated by one atom is absorbed by another \cite{Mandel:95}. This
effect was already observed in 1970 with atoms close to a metallic mirror 
\cite{Drexhage:70}: the mirror image of one atom behaves as a second
radiator in antiphase with the first. Secondly, the DFS conditions given in
Section \ref{DFS-cond} are far more general than the Dicke states. This will
become even more apparent in our discussion on DF subsystems, in Section \ref
{subsystems}. Within the formal analysis provided by the DFS conditions the
Dicke states appear naturally as those states that are isolated against any
interactions with the radiation field.

To further emphasize the distinction between Dicke states and DFS theory, we
consider an interesting extension of these collective atomic states, that
results when one incorporates the single field mode into the system, and
distinguishes this from an external environment of other radiation modes.
This is the situation relevant to a set of $N$ atoms interacting with a
single mode in a high-finesse optical cavity, where this single cavity mode
couples to a free radiation field:\ the prototypical situation in cavity-QED 
\cite{Berman:94}. The Hamiltonian is now
\begin{equation}
\mathbf{H}=\sum_{i=1}^{N}\left( g_{i}^{+}\sigma _{i}^{+}\otimes
b+g_{i}^{-}\sigma _{i}^{-}\otimes b^{\dagger }\right) +\sum_{\lambda }\left(
s_{\lambda }^{-}a_{\lambda }b^{\dagger }+s_{\lambda }^{+}a_{\lambda
}^{\dagger }b\right)   \label{eq:cavity_ham}
\end{equation}
and the system-environment interaction consists only of the second sum,
since the atom-cavity mode interaction is now within the expanded system of
atom plus cavity mode. In this case we find a single DFS condition, $b|\psi
\rangle |n\rangle \propto |\psi \rangle |n\rangle $ where $|\psi \rangle $
is the collective atomic state $|n\rangle $ is the cavity number state. This
condition implies that $b|n\rangle =0$, \textit{i.e.}, the cavity mode must
be empty, but does not otherwise place any restrictions on the atomic
component of the system state. It is only when one imposes a further
requirement that the system Hamiltonian not cause system states to evolve
outside the DFS that one arrives at specific atomic states. In particular,
this requirement tells us that both $\mathbf{S}_{+}b$ and $\mathbf{S}
_{-}b^{\dagger }$ act on $|\psi \rangle |0\rangle $ to produce at most
another state $|\psi ^{\prime }\rangle |0\rangle $. Since $b|0\rangle =0$,
the first condition is trivially satisfied, while the second condition
reduces to $\mathbf{S}_{-}|\psi \rangle =0$ because $b^{\dagger }|0\rangle
=|1\rangle $ which would take the joint state outside the DFS. Thus the
requirement for a DFS of the coupled $N$ atom/cavity system is simply $
\mathbf{S}_{-}|\psi \rangle =0$, with a zero cavity occupation number. The
atomic component of this DFS state is equal to the lowest Dicke state $
|J,-J\rangle $, which constitute the ground
state for a given cooperation number $J$ \cite{Mandel:95}. For $N=2$ we find a two-dimensional DFS given by the atomic states $
|J=1,m=-1\rangle =|00\rangle $ and $|J=0,m=0\rangle =(1/\sqrt{2})[|10\rangle
-|01\rangle ]$ (combined of course with the zero cavity mode state $
|n=0\rangle $). The corresponding DFS for higher values of $N$ may be
readily generated using angular momentum algebra, as described in \cite{Mandel:95}. The dimensionality of these atom/cavity DFS differ from that of
collective decoherence given above [Eq.~(\ref{eq:dim_cd})], being given by
the expression
\begin{equation}
\dim [\mathrm{DFS}(N_{J})]=\frac{(2J+1)N!}{(N/2+J+1)!(N/2-J)!}
\label{eq:dim_cd_new}
\end{equation}
which reduces to Eq.~(\ref{eq:dim_cd}) when $J=0$. Proposals for employing
qubits encoded into these atom/cavity DFSs have been made recently by Beige 
\textit{et al.}, including discussion of performing quantum computation on
the states using the quantum Zeno effect \cite{Beige:99,Beige:00}.

\subsection{Multiple Qubit Errors}
\label{MQE}

The collective decoherence example assumes that all qubits couple to the
bath in a symmetric manner, which is the consequence of full permutational
symmetry. However, the symmetry that gives rise to a DFS can be less strict.
Consider the case of $N=4$ qubits which can undergo bit flips $\sigma
_{j}^{x}$: $|0\rangle _{j}\longleftrightarrow |1\rangle _{j}$.
Suppose that the qubits are arranged in pairs, such that the bit flip error
affects at least two at a time. Specifically, let the allowed errors
be $Q_{X}(4)=\{I,\sigma _{1}^{x}\sigma _{2}^{x},\sigma _{3}^{x}\sigma
_{4}^{x},\sigma _{1}^{x}\sigma _{2}^{x}\sigma _{3}^{x}\sigma _{4}^{x}\}$,
where $I$ is the identity operator (no error). Note that the error $\sigma
_{2}^{x}\sigma _{3}^{x}$ is assumed not to happen. A general error in this
model is a Kraus operator formed by an arbitrary linear combination of the
elements of (the Abelian group) $Q_{X}(4)$. It is then simple to verify that
the encoding 
\begin{eqnarray*}
|0_{L}0_{L}\rangle  &=&\frac{1}{2}\left( |0000\rangle +|1100\rangle
+|0011\rangle +|1111\rangle \right)  \\
|0_{L}1_{L}\rangle  &=&\frac{1}{2}\left( |0001\rangle +|1101\rangle
+|0010\rangle +|1110\rangle \right)  \\
|1_{L}0_{L}\rangle  &=&\frac{1}{2}\left( |0100\rangle +|1000\rangle
+|0111\rangle +|1011\rangle \right)  \\
|1_{L}1_{L}\rangle  &=&\frac{1}{2}\left( |1001\rangle +|0101\rangle
+|1010\rangle +|0110\rangle \right) 
\end{eqnarray*}
is a 4-dimensional DFS (denoted $DFS_{Q_{X}(4)}$), thus encoding two logical
qubits. To see this, check that all states in $DFS_{Q_{X}(4)}$ are
eigenstates with the same eigenvalue $+1$ of any linear combination of
operators over $Q_{X}(4)$. This means that the action of the errors in $
Q_{X}(4)$ is to leave the subspace $DFS_{Q_{X}(4)}$ invariant, so that this
subspace is DF, in accordance with Theorem \ref{th:MQE-DFS} and the more
general Theorem \ref{th:Kraus-DFS}. The latter theorem, however, only
requires that the Kraus operators have an identical unitary representation
on the DFS, which means that eigenvalues other than $+1$ are possible in
principle. Indeed, it is simple to check that the state $(|0000\rangle
-|1100\rangle +|0011\rangle -|1111\rangle )/2$ belongs to another $4$
-dimensional DFS, characterized by the eigenvalue $-1$. In fact the entire $
16$-dimensional Hilbert space splits up into four $4$-dimensional DFSs, two
with eigenvalue $+1$, two with eigenvalue $-1$.

In order to generalize this result, note that the group $Q_{X}(4)$ is
generated (under multiplication) by $\sigma _{1}^{x}\sigma _{2}^{x}$ and $
\sigma _{3}^{x}\sigma _{4}^{x}$; clearly the group $Q_{X}(N)$ generated by $
\{\sigma _{2j-1}^{x}\sigma _{2j}^{x}\}_{j=1}^{N/2}$ also supports a DFS with
the same defining property, namely that the DF states are simultaneous
eigenstates of all $Q_{X}(N)$ generators.

Even more generally, the type of DFS considered here is an example of a 
\emph{DFS for multiple qubit errors}. The general theory is worked out in \cite{Lidar:00a}. The error model can be stated abstractly as follows:
Starting from the Pauli group on $N$ qubits (the group of $N$-fold tensor
products of the Pauli matrices $\{\sigma _{x},\sigma _{y},\sigma _{z}\}$),
look for \emph{Abelian subgroups}. Every such subgroup defines an error
model, in the sense that linear combinations of elements of the subgroup
form Kraus operators. The physical motivation for such subgroups is: (i) The
case in which either only higher order (\emph{multiple qubit}) errors occur,
such that first-order terms of the form $I\otimes \cdots \otimes I\otimes
\sigma _{i}^{\alpha }\otimes I\otimes \cdots \otimes I$ are absent in the
Pauli-group expansion of the Kraus operators, or (ii) only errors of \emph{one} kind, either $\sigma ^{x}$, $\sigma ^{y}$, or $\sigma ^{z}$ take place.
Case (i) would imply that there are certain cancellations involving bath
matrix element terms such that first-order system operators are absent in
the expansion of Eq.~(\ref{eq:Amunu}). While this would be a rather
non-generic situation, involving a very special \textquotedblleft
friendly''\ bath, it turns out that \emph{such conditions
can be actively generated from arbitrary baths} using fast and strong,
dynamical decoupling pulses \cite{WuLidar:01b}. Alternatively, in case (i)
the system-bath Hamiltonian does not have the general form of Eq.~(\ref{eq:H_I}), but rather involves only second order terms such as $\sigma
_{i}^{\alpha }\otimes \sigma _{j}^{\beta }$ (identity on all the rest). Case
(ii) is applicable in, e.g., the case of pure phase damping (relevant to NMR 
\cite{Slichter:book}) and optical lattices using cold controlled collisions 
\cite{Jaksch:99}), where $\sigma ^{z}$ errors are dominant. When either
condition (i) or (ii) above is satisfied, the subgroup is Abelian, which is
a necessary condition for it to have one-dimensional irreps. This, in turn,
is a necessary and sufficient condition for a DFS by Theorem \ref{th:Kraus-DFS}. Thus, to every such multiple qubit error model there
corresponds a DFS, which is found by studying the one-dimensional irreps of
the subgroup over $N$ qubits. The DFS basis states are states that transform
according to a one-dimensional irrep with fixed eigenvalue. The dimension of
such a DFS therefore turns out to be the multiplicity of the one-dimensional
irreps of the subgroup. The result seen above for $Q_{X}(4)$ is general: the
entire Hilbert space splits up into equi-dimensional, independent DFSs \cite{Lidar:00a}. It is also possible to perform universal, fault tolerant
quantum computation on these DFSs, by using a hybrid DFS-QECC method \cite{Lidar:00b}. An experimental magnetic resonance realization of DF quantum
computation, implementing Grover's algorithm in the presence of
multiple-qubit errors, has recently been completed \cite{Ollerenshaw:02}.

\section{Decoherence-Free (Noiseless) Subsystems}

\label{subsystems}

As long as we restrict our attention to encoding quantum information in a
DF manner over a \emph{subspace} then Eqs.~(\ref{eq:DFS-cond}),(\ref{eq:DFD}) provide necessary and sufficient conditions for the
existence of such DFSs. The notion of a subspace which remains DF throughout
the evolution of a system is not, however, the most general method for
providing DF encoding of information in a quantum system. Knill, Laflamme,
and Viola \cite{Knill:99a} discovered a method for DF encoding into \emph{subsystems} instead of into subspaces. Zanardi soon thereafter realized that
this opened the door to a unification of many of the previously seemingly
unrelated ideas for reducing decoherence \cite{Zanardi:99d}. Related, less
general results were independently derived in \cite{DeFilippo:00,Yang:01}. Kempe \textit{et al.} developed a general theory of
universal quantum computation based on the DF subsystems concept \cite{Kempe:00}.

\subsection{Formal Theory}

We first introduce a formal characterization of DF subsystems. This is most
easily done in the Hamiltonian formulation of decoherence. Let $\mathcal{A}$
denote the associative algebra (i.e., the algebra of all polynomials)
generated by the system Hamiltonian $\mathbf{H}_{S}$ and the system
components of the interaction Hamiltonian, the $\mathbf{S}_{\alpha }$'s of
Eq.~(\ref{eq:H_I}). We assume that $\mathbf{H}_{S}$ is expressed in terms of
the $\mathbf{S}_{\alpha }$'s as well (if needed we can always enlarge this
set), and that the identity operator is included as $\mathbf{S}_{0}=\mathbf{I
}_{S}$ and $\mathbf{B}_{0}=\mathbf{I}_{B}$. This will have no observable
consequence but allows for the use of an important representation theorem.
Because the Hamiltonian is Hermitian the $\mathbf{S}_{\alpha }$'s must be
closed under Hermitian conjugation: $\mathcal{A}$ is a ``$
\dagger $-closed'' operator algebra. A basic theorem of
such operator algebras which include the identity operator states that, in
general, $\mathcal{A}$ will be a reducible subalgebra of the full algebra of
operators on the system Hilbert space $\mathcal{H}_{S}$ \cite{Landsman:98a}.
This means that the algebra is isomorphic to a direct sum of $d_{J}\times
d_{J}$ complex matrix algebras $\mathcal{M}(d_{J})$, each with multiplicity $
n_{J}$: 
\begin{equation}
\mathcal{A}\cong \bigoplus_{J\in \mathcal{J}}\mathbf{I}_{n_{J}}\otimes 
\mathcal{M}(d_{J}).  \label{eq:repthm}
\end{equation}
Here $\mathcal{J}$ is a finite set labeling the irreducible components of $
\mathcal{A}$.

The structure implied by Eq.~(\ref{eq:repthm}) is illustrated schematically
as follows, for some system operator $\mathbf{S}_{\alpha }$: 
\begin{equation}
\mathbf{S}_{\alpha }=\left[ 
\begin{tabular}{cccc}
\cline{1-1}
\multicolumn{1}{|c}{$J=1$} & \multicolumn{1}{|c}{} &  &  \\ \cline{1-2}
& \multicolumn{1}{|c}{$J=2$} & \multicolumn{1}{|c}{} &  \\ \cline{2-2}
&  & $\ddots $ &  \\ \cline{4-4}
&  &  & \multicolumn{1}{|c|}{$J=|\mathcal{J}|$} \\ \cline{4-4}
\end{tabular}
\ \right] 
\end{equation}
In this block diagonal matrix representation, a typical block with given $J$, may have a further block diagonal structure, 
\begin{equation}
\left[ 
\begin{tabular}{cccccccccc}
\cline{1-3}
\multicolumn{1}{|c}{} &  &  & \multicolumn{1}{|c}{} &  &  &  &  &  &  \\ 
\multicolumn{1}{|c}{} & $M_{\alpha }$ &  & \multicolumn{1}{|c}{} &  &  &  & 
&  &  \\ 
\multicolumn{1}{|c}{} &  &  & \multicolumn{1}{|c}{} &  &  & $\mu $ &  &  & 
\\ \cline{1-6}
&  &  & \multicolumn{1}{|c}{} &  &  & \multicolumn{1}{|c}{$0$} &  &  &  \\ 
&  &  & \multicolumn{1}{|c}{} & $M_{\alpha }$ &  & \multicolumn{1}{|c}{$
\vdots $} &  &  &  \\ 
&  &  & \multicolumn{1}{|c}{} &  &  & \multicolumn{1}{|c}{$d_{J}-1$} &  &  & 
\\ \cline{4-6}
&  & $\mu ^{\prime }:$ & $0$ & $\cdots $ & $d_{J}-1$ & $\ddots $ &  &  &  \\ 
\cline{8-9}
&  &  &  &  &  &  & \multicolumn{1}{|c}{} &  & \multicolumn{1}{|c}{} \\ 
&  &  &  &  &  &  & \multicolumn{1}{|c}{} & $M_{\alpha }$ & 
\multicolumn{1}{|c}{} \\ 
&  &  &  &  &  &  & \multicolumn{1}{|c}{} &  & \multicolumn{1}{|c}{} \\ 
\cline{8-9}
\end{tabular}
\ \right] 
\begin{tabular}{c}
\\ 
$\lambda =0$ \\ 
\\ 
\\ 
\\ 
$\lambda =1$ \\ 
\\ 
\\ 
\\ 
\\ 
\\ 
\end{tabular}
\end{equation}
Here $\lambda $ labels the different degenerate sub-blocks, and $\mu $
labels the states inside each sub-block. Associated with this decomposition
of the algebra $\mathcal{A}$ is the decomposition over the system Hilbert
space:
\begin{equation}
\mathcal{H}_{S}=\sum_{J\in \mathcal{J}}\CC^{n_{J}}\otimes \CC
^{d_{J}}.  \label{eq:repspc}
\end{equation}
DF subsystems are defined as the situation in which
information is encoded in a single component $\CC^{n_{J}}$ of
Eq.~(\ref{eq:repspc}) (thus the dimension of the DF subsystem
is $n_{J}$). \emph{The decomposition in Eq.~(\ref{eq:repthm}) reveals that
information encoded in such a subsystem will always be affected as identity
on the component} $\CC^{n_{J}}$, \emph{and thus this information will
  not decohere.} This deep fact is the basis for the theory of DF subsystems.

Several general observations are in order: (i) The tensor product structure which gives rise to the name subsystem in Eq.~(\ref{eq:repthm}) is a tensor product over a direct sum, and therefore will not
in general correspond to the natural tensor product of (spatially
distinguishable) qubits; (ii) The
subsystem nature of the decoherence implies that the information should be
encoded in a separable way: over the tensor structure of Eq.~(\ref{eq:repspc}) the density matrix should split into two valid density matrices: $\mathbf{
\rho }_{S}(0)=\mathbf{\rho }\otimes \mathbf{\gamma }$ where $\mathbf{\rho }$
is the DF subsystem and $\mathbf{\gamma }$ is the
corresponding component of the density matrix which \emph{does} decohere;
(iii) Not all of the subsystems in the different irreducible representations
can be simultaneously used: (phase) decoherence will occur between the
different irreducible components of the Hilbert space labeled by $J\in 
\mathcal{J}$. For this reason, from now on we restrict our attention to the
subsystem defined by a \emph{given} $J$.

DF subspaces are now easily connected to DF subsystems. DF subspaces
correspond to those DF subsystems possessing \emph{one-dimensional}
irreducible matrix algebras: $\mathcal{M}(1)$. The multiplicity of these
one-dimensional irreducible algebras is the dimension of the DF subspaces.
In fact it is easy to see how the DF subsystems arise out of a non-commuting
generalization of the DF subspace conditions. Let $\{|\lambda _{\mu }\rangle\}$, $1\leq \lambda \leq n_{J}$ and $1\leq \mu \leq d_{J}$, denote a
subsystem of $\mathcal{H}_{S}$ with given $J$. Then the condition for the
existence of an irreducible decomposition as in Eq.~(\ref{eq:repthm}) is 
\begin{equation}
\mathbf{S}_{\alpha }|\lambda _{\mu }\rangle =\sum_{\mu ^{\prime}=1}^{d_{J}}M_{\mu \mu ^{\prime },\alpha }|\lambda _{\mu ^{\prime }}\rangle ,
\label{eq:subcon}
\end{equation}
for all $\mathbf{S}_{\alpha }$, $\lambda $ and $\mu $. Note that $M_{\mu
\mu ^{\prime },\alpha }$ is \emph{not} dependent on $\lambda $, in the same
way that $c_{\alpha }$ in Eq.~(\ref{eq:DFS-cond}) is the same for all
states $|
\tilde{k}\rangle $ (there $\mu =1$ and fixed). Thus for a fixed $\lambda $, the subsystem spanned by $|\lambda _{\mu }\rangle $ is acted upon in some
nontrivial way. However, because $M_{\mu \mu ^{\prime },\alpha }$ is not
dependent on $\lambda $, each subsystem defined by a fixed $\mu $ and running
over $\lambda $ is acted upon in an \emph{identical manner} by the
decoherence process.

The generalization of the Lindblad semigroup formulation of the DF subspace
condition to the corresponding master equation for a DF subsystem is not
trivial. This is because the $\mathbf{F}_{\alpha }$ operators in Eq.~(\ref{eq:mastereq}) are not required to be closed under conjugation. The
representation theorem Eq.~(\ref{eq:repthm}) is hence not directly
applicable. It can be shown, however, that the master equation analog of
Eq.~(\ref{eq:subcon}) is:$\ \mathbf{F}_{\alpha }|\lambda _{\mu }\rangle
=\sum_{\mu ^{\prime }=1}^{d_{J}}M_{\mu \mu ^{\prime },\alpha }|\lambda _{\mu
^{\prime }}\rangle $, and provides a necessary and sufficient condition for
the preservation of DF subsystems. For details the reader is referred to 
\cite[p.5]{Kempe:00}.

As noted before, the subsystems concept is crucial in establishing a general
theory of universal, fault tolerant quantum computation over DFSs. To this
end it is necessary to introduce the \emph{commutant} $\mathcal{A}^{\prime }$
of $\mathcal{A}$. This is the set of operators which commutes with the
algebra $\mathcal{A}$, $\mathcal{A}^{\prime }=\left\{ \mathbf{X}:[\mathbf{X},
\mathbf{A}]=0,\forall \mathbf{A}\in \mathcal{A}\right\} $. They also form a $
\dagger $-closed algebra, which is reducible to 
\begin{equation}
\mathcal{A}^{\prime }=\bigoplus_{J\in \mathcal{J}}\mathcal{M}(n_{J})\otimes 
\mathbf{I}_{d_{J}}  \label{eq:commutant}
\end{equation}
over the same basis as $\mathcal{A}$ in Eq.~(\ref{eq:repthm}). Comparing to
Eq.~(\ref{eq:repthm}) it is clear that non-trivial operators in $\mathcal{A}
^{\prime }$ are exactly the transformations that preserve the DFS \cite{Zanardi:99a}. As shown in detail in \cite{Kempe:00}, this allows for a
constructive proof that the Heisenberg exchange interaction is universal
over all DFS encodings for collective decoherence.

\subsection{Examples}

\begin{figure}
\centering
\includegraphics[height=15cm,angle=270]{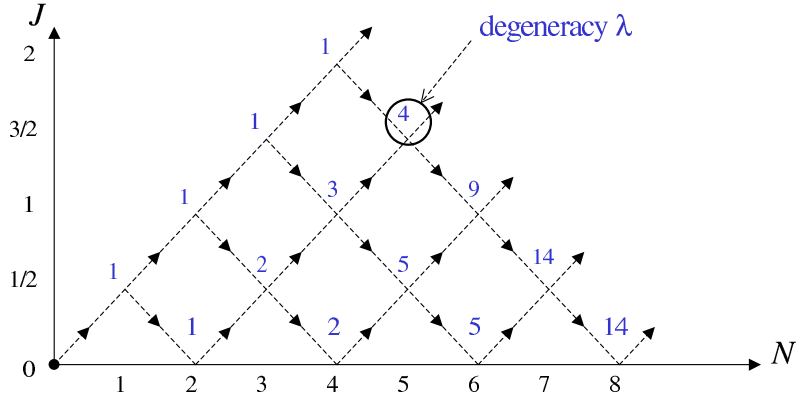}
\vspace{-5cm}
\caption{Partitioning of the Hilbert space of $N$ spin-$1/2$ particles
into DF subspaces (nodes on the $N$-axis) and subsystems (nodes above
the $N$-axis). Integer above a node represents
the number of paths leading from the origin to that node, i.e., the
degeneracy quantum number $\lambda$. See text for full details.}
\label{fig:DFS}
\end{figure}

To illustrate the DF states that emerge from the subsystems formulation it
is convenient to use once more the collective decoherence model.
As will be recalled from Section \ref{spin-boson}, ``strong'' collective decoherence on $N$
qubits is characterized by the three system operators $\mathbf{S}_{+}$, $
\mathbf{S}_{-}$ and $\mathbf{S}_{z}$. These operators form a representation
of the semisimple Lie algebra $sl(2)$. The algebra $\mathcal{A}$ generated
by these operators can be decomposed as\footnote{
Note that as a complex algebra $\{\mathbf{S}_{+},\mathbf{S}_{-},\mathbf{S}
_{z}\}$ span all of $gl(2)$, not just $sl(2)$.} 
\begin{equation}
\mathcal{A}\cong \bigoplus_{J=0(1/2)}^{N/2}\mathbf{I}_{n_{J}}\otimes gl(2J+1)
\label{eq:Agl2}
\end{equation}
where $J$ labels the total angular momentum of the corresponding Hilbert
space decomposition (and hence the $0$ or $1/2$ depending on whether $N$ is
even or odd respectively) and $gl(2J+1)$ is the general linear algebra
acting on a space of size $2J+1$. The resulting decomposition of the system
Hilbert space 
\begin{equation}
\mathcal{H}_{S}\cong \bigoplus_{J=0(1/2)}^{N/2}\CC_{n_{J}}\otimes 
\CC_{2J+1}  \label{eq:Hsdecomp}
\end{equation}
is exactly the reduction of the Hilbert states into different Dicke states 
\cite{Dicke:54,Mandel:95}. The degeneracy for each $J$ is therefore given by
Eq.~(\ref{eq:dim_cd_new}). Eq.~(\ref{eq:Agl2}) shows that given $J$, a state 
$|J,\lambda ,\mu \rangle $ is acted upon as identity on its $\lambda $
component. Thus a DFS is defined by fixing $J$ and $\mu $, where $\mu $ can
now be naturally interpreted as the $z$-component of the projection of the
total angular momentum. As we will show shortly, $\lambda $ corresponds to
the degeneracy of paths leading to a given node $(N,J)$ on the diagram of
Fig.~(\ref{fig:DFS}). In the strong collective decoherence case, we shall
denote the $N$-qubit DFS labeled by a particular angular momentum $J$, by $
DFS_{N}(J)$.

The DFSs corresponding to the different $J$ values for a given $N$ can be
computed using standard methods for the addition of angular momentum. We use
the convention that $|1\rangle $ represents a $|j=1/2,m_{j}=1/2\rangle $
particle and $|0\rangle $ represents a $|j=1/2,m_{j}=-1/2\rangle $ particle
in this decomposition although, of course, one should be careful to treat
this labeling as strictly symbolic and not related to the physical angular
momentum of the particles.

The smallest $N$ which supports a DFS and encodes at least a qubit of
information is $N=3$ \cite{Knill:99a,DeFilippo:00,Yang:01}. In this case
there are two possible values of the total angular momentum: $J=3/2$ or $
J=1/2$. The four $J=3/2$ states $|J,\lambda ,\mu \rangle =|3/2,0,\mu \rangle 
$ ($\mu =m_{J}=\pm 3/2,\pm 1/2$) are singly degenerate; the $J=1/2$ states
have degeneracy $2$. They can be constructed by either adding a $J_{12}=1$
(triplet) or a $J_{12}=0$ (singlet) state to a $J_{3}=1/2$ state. These two
possible methods of adding the angular momentum to obtain a $J=1/2$ state
are exactly the degeneracy of the algebra. The four $J=1/2$ states are: 
\begin{eqnarray*}
&&
\begin{array}{l}
|{\frac{1}{2}},0,{\frac{1}{2}}\rangle ={\frac{1}{\sqrt{2}}}\left( |010\rangle -|100\rangle
\right)  \\ 
|{\frac{1}{2}},0,-{\frac{1}{2}}\rangle ={\frac{1}{\sqrt{2}}}\left( |011\rangle -|101\rangle
\right) 
\end{array}
\\
&&
\begin{array}{l}
|{\frac{1}{2}},1,0\rangle ={\frac{1}{\sqrt{6}}}\left( -2|001\rangle
+|010\rangle +|100\rangle \right)  \\ 
|{\frac{1}{2}},1,1\rangle ={\frac{1}{\sqrt{6}}}\left( 2|110\rangle
-|101\rangle -|011\rangle \right) 
\end{array}
\end{eqnarray*}
where on the left we used the $|J,\lambda ,\mu \rangle $
notation and on the right the states are expanded
in terms the single-particle $|j=1/2,m_{j}=\pm 1/2\rangle $ basis using
Clebsch-Gordan coefficients. The DF qubit can now be written compactly as 
\begin{eqnarray}
|0_{L}\rangle  &=&\alpha |1/2,0,1/2\rangle +\beta |1/2,0,-1/2\rangle \nonumber
\\
|1_{L}\rangle  &=&\alpha |1/2,1,1/2\rangle +\beta |1/2,1,-1/2\rangle 
\label{eq:3dfs}
\end{eqnarray}
where $|\alpha |^{2}+|\beta |^{2}=1$, i.e., the encoding is into the
degeneracy of the two $J=1/2$ subsystems. By the the decomposition of Eqs.~(\ref{eq:Agl2}),(\ref{eq:Hsdecomp}) it is clear that collective errors can
change the $\alpha ,\beta $ coefficients, but have the same effect on the $
|0_{L}\rangle ,|1_{L}\rangle $ states, which is why this encoding is a DFS.

The smallest DF subspace (as opposed to subsystem) supporting a full encoded
qubit comes about for $n=4$. Subspaces for the strong\ collective
decoherence mechanism correspond to the degeneracy of the zero total angular
momentum eigenstates (there are also two DF subsystems with degeneracy $1$
and $3$). As we already noted in Section \ref{spin-boson} this subspace is
spanned by the states: 
\begin{eqnarray}
|0_{L}\rangle  &=&|0,0,0\rangle =|0,0\rangle \otimes |0,0\rangle ={\frac{1}{2
}}|(|01\rangle -|10\rangle )(|01\rangle -|10\rangle )  \nonumber \\
|1_{L}\rangle  &=&|0,1,0\rangle ={\frac{1}{\sqrt{3}}}(|1,1\rangle \otimes
|1,-1\rangle -|1,0\rangle \otimes |1,0\rangle +|1,-1\rangle \otimes
|1,1\rangle )  \nonumber \\
&=&{\frac{1}{\sqrt{12}}}(2|0011\rangle +2|1100\rangle -|0101\rangle
-|1010\rangle -|0110\rangle -|1001\rangle ).  \label{eq:4dfs}
\end{eqnarray}
The notation is the same as in Eq.~(\ref{eq:3dfs}), except that in the
third equality we have used the notation $|J_{12},m_{J_{12}}\rangle \otimes
|J_{34},m_{J_{34}}\rangle $ which makes it easy to see how the angular
momentum is added.

There is a simple graphical interpretation of these results, shown in Fig.~(\ref{fig:DFS}). Starting from the origin (no spins) one starts adding spin-$
1/2$ particles. The first particle ($N=1$)\ has total angular momentum $J=1/2$. This single particle has the trivial $DFS_{1}(1/2)$. When the second
particle is added one can form either a singlet ($J=0$) by subtracting $1/2$ or form a triplet by adding $1/2$. In this manner the subspace $DFS_{2}(0)$
and subsystem $DFS_{2}(1)$ are formed, containing a single state ($\mu =m_J=0$)
and three states ($\mu =m_J=-1,0,1$), respectively. The first interesting DFS is
formed when $N=3$, since now there are two paths, $\lambda =0,1$, in Fig.~(\ref{fig:DFS}) leading to the same total $J=1/2$. These two paths correspond
exactly to the $|0_{L}\rangle $ and $|1_{L}\rangle $ states of Eq.~(\ref{eq:3dfs}), and define the subsystem $DFS_{3}(1/2)$, that encodes a single
DF qubit. For $N=4$ spins there is the DF subspace $DFS_{4}(0)$ and also a
three-dimensional DF subsystem $DFS_{4}(1)$. Clearly, the DF subspaces are
simply all the nodes that lie on the $N$-axis in Fig.~(\ref{fig:DFS}), while
the DF subsystems are all the nodes above this axis. Interpreted in this
manner, Fig.~(\ref{fig:DFS}) is a partitioning of the entire system Hilbert
space into disjoint DFSs. The examples presented here have been worked out
in far greater detail in \cite{Kempe:00,Bacon:thesis,Kempe:thesis},
where it is further shown how to 
construct universal quantum computation operations using the same formalism.

\section{Protection against additional decoherence sources}
\label{robustness}

It is clear from the above summary of DF subspaces and subsystems that
they will afford complete protection against
the specific errors that they are designed to eliminate. However, in
general it is not possible to protect agains all kinds of errors and
one is motivated to then consider how to add protection of DFS
encodings agains additional sources of decoherence. Several approaches
have been taken to address this issue.  It was shown early on that DFSs are remarkably stable to additional sources of decoherence,
providing a useful point of departure for more general schemes to
protect fragile quantum information.  The
first study to consider the effects of perturbations breaking the
symmetry giving rise to a DFS was Ref.~\cite{Lidar:PRL98}, within the
Markovian semigroup formulation. It was
shown that if an additional set of Lindblad operators are added to the
Lindblad equation (\ref{eq:mastereq}), parametrized by a coupling
strength $\epsilon$, then the ``mixed state fidelity'' $f_\epsilon(t) \equiv
{\rm Tr}[\rho_{\rm id}(t)\rho_{\epsilon}(t)]$, i.e., the overlap
  between the ideal (decoherence-free) evolution $\rho_{\rm id}(t)$
  and the perturbed evolution $\rho_{\epsilon}(t)$, decays as
  $1-O(\epsilon t^2)$. This
conclusion was significantly strengthened in Ref.~\cite{Bacon:99},
where a
stability analysis was made within both the semigroup and OSR
formulations. It was shown that in both formulations states in DFSs are stable to
symmetry-breaking perturbations to second
order in the strength of the perturbation: $f_\epsilon(t) =
1-O(\epsilon^2 t^2)$, so the entire $O(\epsilon)$ contribution
vanishes to all orders in time.  This indicates
that the DFSs are well-suited for applications to quantum memory, a
result that has been recently been taken advantage of in ion
traps \cite{Kielpinski:01,Kielpinski:02,Brown:02,LidarWu:02}.  Errors
that do occur can be dealt with by a number of methods: (i) Additional
encoding \cite{Lidar:PRL99,Lidar:PRA00Exchange,Alber:01,Alber:01a,KhodjastehLidar:02}, (ii) Implementing
error-detection circuits \cite{Kempe:01}; (iii) Hamiltonian
engineering to suppress or eliminate further errors \cite{Bacon:01}; (iv) Combination with the dynamical
decoupling technique \cite{WuLidar:01b,LidarWu:02,WuByrdLidar:02,ByrdLidar:01a,ByrdLidar:02a,Viola:01a}.  
We briefly summarize instances of these four approaches here. 

In the collective
decoherence DFS, independent single qubit errors can be shown to cause
either independent errors acting on the encoded qubit states, or
leakage of the DFS encoded qubit states to states lying outside the
DFS \cite{Lidar:PRL99}.  Both of these types of errors may be corrected
by concatenating the DFS encoding with a standard quantum
error-correcting code that corrects independent single qubit errors.
Thus, the four qubit DFS encoding for collective decoherence was
concatenated with the five-qubit QECC in Ref.~\cite{Lidar:PRL99} to
produce a 20-qubit encoding that provides protection against both
collective and independent errors on the physical qubits.  
Concatenation thus provides one way to correct leakage 
errors that take the encoded quantum information out of the DFS.

Another approach
is to implement a leakage-correction
circuit \cite{Preskill:97a}.  An example of this for the
three-qubit encoding into a subsystem protected against collective
decoherence is given in Ref.~\cite{Kempe:01}.  Such circuits may
be readily incorporated into schemes for implementing fault-tolerant
computation such as in Refs.~\cite{Bacon:99a,Kempe:00}.  This
particular leakage-correction circuit allows correction without any
measurement, providing a simpler approach than usual in protocols for
fault tolerance.

The third approach to dealing with errors not
specifically protected against by the DFS encoding, namely by Hamiltonian
engineering 
to eliminate additional errors, is very different from the first two.  The idea here is
to supplement the qubit Hamiltonian by additional interactions that
impose an energy spectrum on the DFS states.  The additional
Hamiltonian terms are specifically engineered to result in an energy
spectrum that eliminates or thermally suppresses additional errors.
This approach has been demonstrated in Ref.~\cite{Bacon:01},
using the four-qubit DFS encoding against collective decoherence.
Addition of a specifically designed set of exchange interactions
between the physical qubits causes the qubit tensor product states in
the DFS representation (Section~\ref{spin-boson}) to split into a
spectrum of higher angular momentum states that are separated from the
degenerate ground state that provides the zero angular momentum DFS
encoding.  The effect of this additional Hamiltonian is to transform
all independent single-qubit errors into non-energy conserving errors
that cause excitation out of the ground state encoding.  Thus all
local phase errors on the encoded qubit states have been removed,
while maintaining their protection against collective decoherence,
leading to their designation as ``coherence-preserving'' or
``supercoherent''. The energy-nonconserving errors can be suppressed by
going to low temperatures, or alternatively, used as an
error-detecting code. 

The fourth approach to dealing with
additional errors combines DFS encoding with the method of strong and
fast dynamical decoupling pulses proposed by Viola and Lloyd in
Ref.~\cite{Viola:98}. The first concrete example of this approach was
given by Wu and Lidar in Ref.~\cite{WuLidar:01b}, where it was shown that dynamical
decoupling pulses can create the conditions for both the
multiple-qubit and collective decoherence error models, starting from
arbitrary system-bath interactions. In the latter error model, this
can be done by rapidly pulsing the Heisenberg exchange interaction, so
the scheme is fully compatible with universal quantum computation on
the resulting DFS. However, even after the creation of the appropriate
symmetry, leakage from the DFS is still possible. To this end
Refs.~\cite{WuByrdLidar:02,ByrdLidar:01a,ByrdLidar:02a} showed
explicitly how
leakage errors too can be eliminated using dynamical decoupling, and
in the case of collective decoherence, these can again be generated
using Heisenberg interactions. Integration of all these components (DFS protection,
generation of symmetric system-bath interaction, and leakage
elimination), has been discussed in Ref.~\cite{LidarWu:02}.

These studies of effects of errors beyond the specific error models
defining the DFS and its protection show that there are a number of
viable approaches to incorporate DFS encodings into stronger schemes
that provide protection of quantum states at multiple levels.  They
can be used both to further improve the performance of DFS encoded
qubits for quantum memory applications, and also in implementations
of quantum computation with DFS encodings \cite{Kempe:00}.
Concatenation and leakage-detection circuits provide a direct route to
the established framework of fault-tolerance.  Similarly, the use of
Hamiltonian engineering as a design tool to suppress or remove
additional errors relates to the very different approach advocated by
topological quantum computing, where one seeks to optimize the
Hamiltonian structure to eliminate or automatically correct
errors \cite{Kitaev:97,Kitaev:book}. The DFS-dynamical decoupling
combination ties in naturally with several solid state and atomic
physics proposals for quantum computing. These relations deserve further
exploration in the future.

\section{Conclusions}
\label{conclusions}

This review is an introduction to the now
well-developed theory of decoherence-free subspaces and
subsystems (DFSs). The central idea behind DFSs is that decoherence can be avoided by encoding
quantum information into special subspaces that are protected from the
interaction with the environment by virtue of a dynamical
symmetry. This idea has roots that can be traced to work in quantum
optics, the measurement problem, and the general theory of open
quantum systems. However, the appreciation of its importance to
quantum information theory, the development of a rigorous underlying
mathematical framework, and experimental verification, are all very
recent. We have provided here a survey of these
developments, emphasizing the theoretical contributions to the problem
of {\em preservation} of quantum
information, as opposed to the rich and fascinating subject
of its {\em manipulation}. This was illustrated with examples from quantum optics
and general models of interacting quantum systems. It is important to
emphasize that DFSs are only one of several ideas for the preservation
of quantum information, and that full protection of quantum
information is likely to involve a combination at
different levels of several methods. The most notable alternatives are quantum
error correcting codes, geometric and topological encodings of quantum
information,
and dynamical suppression of decoherence via the application of fast
and strong pulses. It is now increasingly apparent that there are
strong unifying features to all of these ideas, in the sense that
there is an underlying algebraic structure from which all the
different methods seem to emerge. Future investigations will no doubt
further clarify this unity, and it is our hope that new
experiments will confirm the usefulness of the DFS concept, in particular
for quantum information processing.

\section*{Acknowledgements}
This material is based on research sponsored by the Defense Advanced
Research Projects Agency under the QuIST program and managed by the Air
Force Research Laboratory (AFOSR), under agreements F49620-01-1-0468 and F30602-01-2-0524. The authors' effort was also supported by the
National Security Agency (NSA) and the Advanced Research and Development Activity (ARDA)
under Army Research Office (ARO) contracts DAAG55-98-1-0371 and DAAD19-00-1-0380.  
D.A.L. further gratefully acknowledges
financial support from the Premier's Research Excellence Award
(Ontario), Photonics Research Ontario, NSERC, and the Connaught Fund. 
K.B.W. thanks the 
Miller Institute for Basic Research in Science for a Miller Research Professorship for 2002-2003.

\end{document}